%% file: main.tex
\documentclass[sigconf]{acmart}

\usepackage{mathtools}
\usepackage{textcomp}
\usepackage{enumerate}
\usepackage[ruled,vlined,linesnumbered]{algorithm2e}
\usepackage{algpseudocode}
\usepackage{booktabs}
\usepackage{makecell}
\usepackage{xparse}
\usepackage{xspace}
\usepackage{multirow}
\usepackage{csvsimple}
\usepackage{pifont}
\usepackage{xcolor}
\usepackage{colortbl}
\usepackage{placeins}
\usepackage[most]{tcolorbox}
\usepackage{float}
\usepackage{dblfloatfix}
\usepackage{longtable}
\usepackage[frozencache,cachedir=.]{minted}
\usepackage{fancyvrb}
\usepackage{wrapfig}
\usepackage{subcaption}
\usepackage{microtype}
\usepackage{enumitem}
\usepackage{breakurl}
\setlist[itemize]{listparindent=\parindent, parsep=0pt, partopsep=0pt}

\AtBeginDocument{%
  }

\algnewcommand{\algorithmicand}{\textbf{ and }}
\algnewcommand{\algorithmicor}{\textbf{ or }}
\algnewcommand{\OR}{\algorithmicor}
\algnewcommand{\AND}{\algorithmicand}
\algnewcommand{\var}{\texttt}
\definecolor{light-gray}{gray}{0.95}
\definecolor{darkblue}{rgb}{0.0, 0.0, 0.55}
\definecolor{darkcandyapplered}{rgb}{0.64, 0.0, 0.0}
\definecolor{mygray}{gray}{0.9}
\definecolor{dkgreen}{rgb}{0,0.6,0}
\definecolor{mauve}{rgb}{0.58,0,0.82}
\definecolor{boxorange}{HTML}{F6B26B}
\definecolor{boxgreen}{HTML}{93C47D}
\definecolor{boxblue}{HTML}{6FA8DC}
\definecolor{lightbg}{HTML}{FFF8ED}

\tcbset{
  fontupper=\ttfamily\footnotesize,
  boxrule=0.5pt,
  arc=6pt,
  coltitle=black,
  left=5pt,
  right=5pt,
  top=5pt,
  bottom=5pt,
  enhanced,
}
\newtcolorbox{instrbox}{
  colback=boxorange!5,
  colframe=boxorange,
}
\newtcolorbox{formatbox}{
  colback=boxgreen!5,
  colframe=boxgreen,
}
\newtcolorbox{exbox}{
  colback=boxblue!5,
  colframe=boxblue,
}

\newcommand{\listingref}[1]{\hyperref[#1]{Listing~\ref*{#1}}}

\def\Snospace~{\S{}}

\newcommand{\sref}[2]{\hyperref[#2]{#1 \ref{#2}}}

\newcommand{\patchagent}{\textsc{PatchAgent}\xspace}
\newcommand{\santopatch}{\textsc{San2Patch}\xspace}
\newcommand{\testname}{\texorpdfstring{$\text{PoC}^+$}{PoC+}\xspace}
\newcommand{\dataset}{\textsc{PVBench}\xspace}

\newcommand{\tinyfootnote}{\fontsize{6.1pt}{7.7pt}\selectfont}

\setcopyright{none}
\acmConference[]{}{}{}
\renewcommand\footnotetextcopyrightpermission[1]{}
\begin{document}

\title{Patch Validation in Automated Vulnerability Repair}

\author{
    Zheng Yu$^{\dagger\S}$,
    Wenxuan Shi$^{\S}$,
    Xinqian Sun$^{\dagger\S}$,
    Zheyun Feng$^{\ddagger}$,
    Meng Xu$^{\dagger}$,
    Xinyu Xing$^{\S}$ \\
    $^{\S}$\textit{Northwestern University}
    $^{\dagger}$\textit{University of Waterloo},
    $^{\ddagger}$\textit{University of New Hampshire} \\
    \textit{\{zheng.yu,wenxuan.shi,xinqian.sun,xinyu.xing\}@northwestern.edu}, \\
    \textit{meng.xu.cs@uwaterloo.ca, Zheyun.Feng@unh.edu}
}

\input{sections/0_abstract}




\renewcommand{\shortauthors}{}
\renewcommand{\shorttitle}{}
\maketitle
\pagestyle{plain}

\input{sections/1_intro}
\input{sections/2_background}

\input{sections/3_motivation}

\input{sections/4_dataset}

\input{sections/5_overestimation}

\input{sections/6_reliability}
\input{sections/7_false_positive}

\input{sections/8_implication}
\input{sections/9_discussion}

\input{sections/10_related}
\input{sections/11_conclusion}


\bibliographystyle{ACM-Reference-Format}
\bibliography{ref}


\input{sections/appendix}

\end{document}

%% file: sections/0_abstract.tex
\begin{abstract}

Automated Vulnerability Repair (AVR) systems, especially those leveraging large language models (LLMs), have demonstrated promising results in patching vulnerabilities---that is, if we trust their patch validation methodology. Ground-truth patches from human developers often come with new tests that not only ensure mitigation of the vulnerability but also encode extra semantics such as root cause location, optimal fix strategy, or subtle coding styles or conventions. And yet, none of the recent AVR systems verify that the auto-generated patches additionally pass these new tests (termed as $\text{PoC}^+$ tests). This is a subtle yet critical omission.

To fill this gap, we constructed a benchmark, $\textrm{PVBench}$, with 209 cases spanning 20 projects. Each case includes basic tests (functional tests before the patch and the PoC exploit) as well as the associated $\text{PoC}^+$ tests. Evaluated on three state-of-the-art AVR systems, we find that over 40\% of patches validated as correct by basic tests fail under $\text{PoC}^+$ testing, revealing substantial overestimation on patch success rates. Analyzing these patches that are falsely labeled as correct, we suggest that AVR tools should improve in three critical areas: root cause analysis, adherence to program specifications, and capturing developer intention.

\end{abstract}

%% file: sections/1_intro.tex
\section{Introduction}

Patch validation---ensuring that a patch effectively addresses security vulnerabilities while preserving functional integrity---is crucial in software development. While traditionally applied to human-written patches ~\cite{wu2023mitigating,le2014patch}, patch validation has become increasingly important for automated vulnerability repair (AVR) systems ~\cite{pearce2023examining,SoKAVR,li2025sok}, where it serves to evaluate auto-generated patches. Hence, the reliability of AVR effectiveness evaluation therefore depends on the accuracy and comprehensiveness of their underlying patch validation methodologies.

In an abstract view,
an AVR tool takes, at minimum,
a vulnerable codebase $C$ and
a proof-of-concept (PoC) input $poc$ that
demonstrates the existence of a vulnerability in $C$,
and produces a patch $\hat{p}$ that is supposed to fix the vulnerability.
Many AVR tools \cite{hong2020saver,xing2024if,gao2021beyond} that achieved good performance in the pre-LLM era focused exclusively on vulnerabilities exhibiting fixed patterns, limiting their generalizability to other vulnerability classes.
%
%
As code generation and completion capabilities advance through machine learning techniques, particularly large language models (LLMs),
researchers are increasingly exploring their application in AVR~\cite{li2025sok,SoKAVR,pearce2023examining}. Several recent studies~\cite{appatch,PatchAgent,SAN2PATCH} have demonstrated promising repair rates 
far exceeding best AVR tools before the LLM era.

Regarding validating the generated patch $\hat{p}$,
our survey of the recent literature reveals three primary methodologies
adopted in previous AVR works, including
\textit{manual comparison} that relies on expert assessment~\cite{xing2024if,appatch,kulsum2024case},
\textit{similarity metrics} that employ similarity-based scoring~\cite{zhou2024out,fu2022vulrepair,chen2022neural}, and
\textit{test suite validation} that executes automated tests to validate patch correctness~\cite{PatchAgent,SAN2PATCH,zhang2024fixing,pearce2023examining}.
Both manual comparison and similarity metrics
compare against a ground-truth patch $p$ which is typically
developer-written and committed into the codebase as the official bugfix.
While manual comparison can provide high accuracy in principle, it lacks scalability for large-scale evaluation ~\cite{ye2021automated,le2019reliability}. Similarity-based metrics offer automation but have been shown to inadequately correlate with actual patch effectiveness~\cite{zhang2024fixing}, as patches achieving high similarity scores may still fail to address the vulnerability.


Test suite-based validation, on the other hand,
does not necessarily require a ground-truth patch $p$ to compare $\hat{p}$ with.
Instead, it assumes that
the codebase $C$ comes with a comprehensive test suite $T$ that
pins down the intended behaviors of the program.
Therefore, if the patched codebase
($\hat{p}$ applied to $C$) passes all tests in $T$ and additionally
mitigates $poc$, then $\hat{p}$ is correct.
%
Perhaps due to its close resemblance to
typical CI/CD pipelines in mature codebases when bugfixes are introduced,
test suite-based validation has emerged as the predominant method in AVR research and practice, adopted by the majority of state-of-the-art tools including \patchagent~\cite{PatchAgent}, \santopatch~\cite{SAN2PATCH}, SWE-Agent~\cite{yang2024sweagent} and others~\cite{shariffdeen2025vulnerability,zheng2025fixing,pearce2023examining,zhang2022program}.


However,
while an AVR tool should never has access to the ground-truth patch $p$
when generating $\hat{p}$,
it does not implies that
we should evaluate the AVR tool completely ignoring $p$.
%
%
In fact,
based on the generally accepted principle that
\emph{new code should have a new test}
we expect new 
tests $t$
associated with the ground-truth patch $p$
in mature open-source projects---effectively,
this means an updated test suite $T'$ in the patched codebase.
So this raises a question:
\textbf{shall we evaluate $\hat{p}$ with $T'$ instead of $T$ and $poc$?}

\begin{table*}[t]\footnotesize
\centering
\caption{Taxonomy of Patch Validation Methodologies Used in Previous AVR Works}

\renewcommand{\arraystretch}{1.4}
\begin{tabular}{p{1.5cm}p{1.2cm}p{4.9cm}p{8.48cm}}
\toprule
\centering\textbf{Method} & \centering\textbf{Category} & \textbf{Papers} & \textbf{Key Characteristics} \\
\midrule
\multirow{4}{1.5cm}{\centering\textbf{Manual Comparison}} & 
    \multirow{1.7}{1.2cm}{\centering Limited Scope} & 
        \textsc{Conch}~\cite{xing2024if}, 
        FixMorph~\cite{shariffdeen2021automated}, 
        SkyPort~\cite{shi2022backporting}, 
        TSBPort~\cite{yang2023enhancing}, 
        PortGPT~\cite{port_gpt}
        
        & 
        Check if patch is semantically equivalent to ground truth;
        The AVR tool is
        constrained to specific bug types
        (e.g., null dereference) or scenarios (e.g., backporting) only. \\
    \cmidrule(lr){2-4}
    & 
    \multirow{1.7}{1.2cm}{\centering General Scope} & 
    \multirow{1.7}{4.8cm}{
            Senx~\cite{huang2019using},
            APPatch~\cite{appatch},
            \textsc{VRpilot}~\cite{kulsum2024case}
        }
    & 
        Check if patch fixes the bug without breaking code functionality;
        The AVR tool
        handles various vulnerability types without constraints. \\
    \midrule
\multirow{1.9}{1.5cm}{\centering\textbf{Similarity Metrics}} & 
    \multirow{1.9}{1.2cm}{\centering Similarity Scoring} & 
    \multirow{1.9}{4.8cm}{
        VulMaster~\cite{zhou2024out},
        VulRepair~\cite{fu2022vulrepair},
        VRepair~\cite{chen2022neural} 
        }
        & 
        Automated evaluation using common code similarity metrics: EM, BLEU-4, and CodeBLEU metrics. \\
    \midrule
\multirow{5}{1.5cm}{\centering\textbf{Test Suite Validation}} & 
    \multirow{1.7}{1.2cm}{\centering PoC-only Testing} & 
        ExtractFix~\cite{gao2021beyond}, CodeRover-S~\cite{zhang2024fixing}, 
        SAVER~\cite{hong2020saver}, 
        VFix~\cite{xu2019vfix}, \textsc{CrashFixer}~\cite{mathai2025crashfixer}
        & 
        Validate patches by executing PoC to confirm vulnerability mitigation; focuses on security-specific validation but may overlook functional correctness. \\
    \cmidrule(lr){2-4}
    & 
    \multirow{2.5}{1.2cm}{\centering Full Testing} & 
        \textsc{CrashRepair}~\cite{shariffdeen2025vulnerability}, 
        CPR~\cite{shariffdeen2021concolic},
        Fix2Fit~\cite{gao2019crash}, \patchagent~\cite{PatchAgent},  \santopatch~\cite{SAN2PATCH}, WilliamT~\cite{zheng2025fixing}, VulnFix~\cite{zhang2022program},
        Zero-Shot~\cite{pearce2023examining}
        & Validate patches by executing both PoC tests and the program's existing functional test suites; ensures patches fix vulnerabilities while maintaining program functionality (test coverage dependent). \\
\bottomrule
\end{tabular}

\label{tab:patch_validation_taxonomy}
\end{table*}

This question seems trivial at first glance,
as it is tempting to assume that $t$ and $poc$ are equally useful
in evaluating $\hat{p}$
(hence $T$ combined with $poc$ is effectively the same as $T'$).
However,
as shown in~\autoref{sec:motivation},
a patch $\hat{p}$ that mitigates $poc$ and passes $T$
does not mean that it fixes the vulnerability in the way intended by the developer,
and such intention can be encoded in the new test $t$ associated with
the patch $p$ from the developers.
This validation gap raises a critical question:
\textbf{does the use of $T$ and $poc$ validation
substantially overestimate the performance of AVR tools
due to their inability to guarantee patch correctness?}

We name the new test $t$
written by developers for the official patch $p$ as \testname in this paper
to honor the fact that
$t$ is often derived from the $poc$ but potentially
encodes more semantics in terms of how developers intend to fix the vulnerability,
and we argue that the auto-generated patch $\hat{p}$ of AVR tools
should be evaluated using $T'$, including the \testname.
We constructed a dataset called \dataset,
comprising 209 cases across 20 open-source projects.
Each case includes both basic tests ($T$ and $poc$) and \testname tests ($t$).
Using \dataset,
we re-evaluate three state-of-the-art AVR tools, 
\patchagent~\cite{PatchAgent}, 
\santopatch~\cite{SAN2PATCH}, and 
SWE-Agent~\cite{zan2025multi,yang2024sweagent}.
%
Our results reveal that over 40\% of patches validated as correct by basic tests failed when evaluated against \testname tests, corresponding to a high false discovery rate in statistical terms. This stark contrast exposes a critical weakness in test suite-based evaluation: commonly used validation methods substantially overestimate AVR tool effectiveness.

To assess whether \testname tests serve as a reliable patch validation method, we manually compared patches that passed \testname tests against developer-written patches. We found that over 70\% of \testname-passing patches achieve semantic equivalence with the developer's patch, demonstrating \testname's effectiveness in capturing developer intent and the inherent repair logic. The remaining patches exhibit issues such as suboptimal complexity or other quality concerns, which we detail in \autoref{sec:true_positive}. Additionally, we analyze patches that fail \testname tests and summarize the failure reasons in \autoref{sec:finding}.

To understand how \testname tests are commonly created by developers (and why they encode more program semantics than the $poc$), we surveyed the \testname tests in \dataset and identified three categories based on their validation mechanisms: \textit{Output Checking}, \textit{Intermediate Checking}, and \textit{Self Checking}. Across all categories, developers transform the original PoC by capturing the expected behavior of the patched program, whether through output comparison, intermediate state assertions, or embedded runtime checks, and encoding these expectations into the test specification.
Unlike a PoC, which only observes whether the program crashes, a \testname test encodes richer program semantics by explicitly specifying the expected correct behavior. This additional semantic information enables more precise patch validation: a patch that merely suppresses a crash without restoring correct functionality will fail the \testname test, whereas it might pass a crash-only PoC check.

In summary, our work makes four main contributions:

\begin{itemize}[noitemsep,nolistsep]
\item We propose \testname tests as an improved patch validation method for evaluating AVR tools and introduce \dataset to investigate their advantages over traditional test suite-based validation. We also describe our validation methodology and the process for producing these tests.

\item We evaluate three state-of-the-art LLM-based AVR systems
on \dataset,
revealing that over 40\% of patches validated as correct by basic tests fail when subjected to \testname tests,
demonstrating substantial overestimation
in current evaluation practices for modern AVR tools.

\item We validate the reliability of \testname tests by comparing patches that pass them against developer patches. Over 70\% achieve semantic equivalence with developer patches, while the remainder exhibit performance issues or other concerns. In other words, \testname does
not fully capture every detail in developers’ intention, but is still an important step forward.

\item We provide a comprehensive analysis of validation inadequacies by systematically categorizing falsely correct
patches into three categories which might help guide APR
tools towards better repair process.

\end{itemize}

To foster further research, we have released the \dataset and evaluation artifacts at \url{https://github.com/cla7aye15I4nd/PVBench}. Additional details are provided in \autoref{sec:open_science}.


%% file: sections/2_background.tex
\section{Background: Patch Validation in AVR}\label{sec:patch-validation-sok}

This section surveys 
  patch validation methods used in previous AVR works
(\autoref{tab:patch_validation_taxonomy}),
examining their practices and potential limitations.



\noindent\textbf{Manual Comparison.} This approach relies on human experts to assess patch quality through direct comparison with developer-provided patches. \textit{Limited scope systems} target specific vulnerability types or constrained scenarios (e.g., \textsc{Conch}~\cite{xing2024if} focuses on null pointer dereferences, while TSBPORT~\cite{yang2023enhancing}, SKYPORT~\cite{shi2022backporting}, PortGPT~\cite{port_gpt}, and \textsc{FixMorph}~\cite{shariffdeen2021automated} address backporting tasks). \textit{General scope systems} such as APPatch~\cite{appatch}, \textsc{VRpilot}~\cite{kulsum2024case}, and Senx~\cite{huang2019using} handle diverse vulnerability types, making evaluation more challenging since correct patches may differ structurally from reference solutions. Unfortunately, none of these works disclose the specific criteria examiners use during comparison, particularly when evaluating functionally correct but structurally different patches.

\emph{Limitations:}
Manual comparison provides arguably the highest level of accuracy,
this approach is typically confined to scenarios with
limited solution spaces.
For instance,
\textsc{Conch}~\cite{xing2024if} focuses specifically on
null dereference vulnerabilities
that require straightforward null check additions,
while three other studies~\cite{yang2023enhancing,shi2022backporting,shariffdeen2021automated} examine backporting scenarios where patches are adapted from mainline to stable versions.
These backporting studies reveal that
most patches require only minimal modifications---%
TSBPORT~\cite{yang2023enhancing} found that
81\% of backported patches involve location or namespace changes only---making
manual verification feasible 
even for backporting tasks on large codebases (e.g., Linux kernel).
For the other works that target general vulnerability repair~\cite{appatch,kulsum2024case,huang2019using},
they all acknowledge that manual review is expensive and typically sample only a subset of cases for evaluation, highlighting the critical need for scalable automated patch validation methods. 

\noindent\textbf{Similarity Metrics.} These approaches employ automated scoring without human intervention, typically combining three metrics: Exact Match (EM)~\cite{rajpurkar2016squad} for binary character-level comparison, BLEU-4~\cite{papineni2002bleu} for token-level n-gram precision, and CodeBLEU~\cite{ren2020codebleu} which extends BLEU with AST-based syntactic and data-flow semantic analysis. Representative systems include VulMaster~\cite{zhou2024out}, VulRepair~\cite{fu2022vulrepair}, and VRepair~\cite{chen2022neural}.

\emph{Limitations:}
While code similarity metrics can be an automated way of measuring patch quality, they often fail to provide reliable indicators of patch effectiveness, as patches with high similarity scores may still fail when executed against PoCs~\cite{zhang2024fixing}. This occurs because even minor syntactic differences—such as incorrect variable names, missing edge case handling, or subtle logic errors—can render functionally similar-looking code completely ineffective. Critical factors like proper variable scoping, correct API usage, and precise conditional logic placement are often overlooked by similarity metrics despite being essential for patch functionality.

\noindent\textbf{Test Suite-Based Validation.} This widely adopted method validates patches through automated test execution. \textit{PoC testing} focuses on exploit validation, exemplified by CodeRover-S~\cite{zhang2024fixing}, ExtractFix~\cite{gao2021beyond}, SAVER~\cite{hong2020saver}, and VFix~\cite{xu2019vfix}, though it may overlook functional correctness. \textit{Comprehensive testing} combines PoC validation with functional test suites in the pre-patched codebase, as adopted by \patchagent~\cite{PatchAgent}, \santopatch~\cite{SAN2PATCH}, \textsc{CrashRepair}~\cite{shariffdeen2025vulnerability}, \textsc{William}~\cite{zheng2025fixing}, Zero-Shot~\cite{pearce2023examining}, \textsc{VulnFix}~\cite{zhang2022program}, CPR~\cite{shariffdeen2021concolic}, and Fix2Fit~\cite{gao2019crash}.

\emph{Limitations:}
Test suite-based validation in AVR remains contentious
due to debating \emph{overfitting} concerns,
where patches may pass tests without addressing root causes.
While early studies argued AVR tools are susceptible to overfitting~\cite{smith2015cure,xiong2018identifying,wang2020automated}, recent work~\cite{petke2024patch,xia2023automated,ouyang2024benchmarking} suggests modern LLM-based AVR tools exhibit less project-specific fitting due to training on diverse codebases. However, existing research fails to explain why patches pass tests yet miss root causes, and largely excludes memory-unsafe languages like C/C++ despite well-studied vulnerability patterns~\cite{song2019sok,yu2024shadowbound,ullah2024llms}.
This knowledge may bias LLMs toward symptom-fixing
rather than addressing root causes
as shown in~\autoref{sec:motivation}.

\noindent\textbf{State-of-the-practice.}
Despite these potential problems,
most modern AVR research continues
to rely on test suite-based validation
to decide patch correctness.
This is likely due to its fully automated nature \cite{SoKAVR,li2025sok} and
its resemblance on
how a manually written patch is applied on a production-grade codebase,
which typically, and minimally, requires that
all new code pushed to production must pass the CI/CD pipeline
\cite{githubci} which runs all existing tests.
However,
test suite-based validation makes a strong assumption on the
comprehensiveness of existing test suite:
the test suite should cover \emph{all} intended behaviors
of the program,
which is unlikely to be true even for mature codebases \cite{ivankovic2019code}.
%
The practical implications of this 
assumption
are illustrated in our motivating example (\autoref{sec:motivation}),
which demonstrates how flawed patches
can bypass current test suites.
Our quantitative analysis in~\autoref{sec:overestimation}
further reveals the extent to which
test suite-based validation influences AVR performance.

%% file: sections/3_motivation.tex
\begin{figure}[t]
    \centering
    \begin{minted}[fontsize=\scriptsize,linenos,breaklines, mathescape,numbersep=5pt,xleftmargin=5pt]{c}
PHP_FUNCTION(range) {
  struct zval *user_start = /* Extract start argument */;
  struct zval *user_end = /* Extract end argument */;
  // Extract type from arguments
  uint8_t start_type = Z_TYPE_P(user_start);
  uint8_t end_type = Z_TYPE_P(user_end);
  /* If the range is given as strings, 
     generate an array of characters. */
  if (start_type >= IS_STRING || end_type >= IS_STRING) {
    // VULNERABLE: condition fails when start_type=5 (IS_DOUBLE)
    // and end_type=7 (IS_ARRAY) because 5+7 = 2*6 (IS_STRING)
    if (start_type + end_type < 2*IS_STRING) {
      // ... handle mixed type inputs and convert to numeric
      goto handle_numeric_inputs;
    }
    // TYPE CONFUSION OCCURS HERE:
    //   When the vulnerable condition fails, we reach this point
    //   with non-string types, but try to access them as strings
    unsigned char low = Z_STRVAL_P(user_start)[0];
    unsigned char high = Z_STRVAL_P(user_end)[0];
    // ... character range generation logic
    return;
  }
handle_numeric_inputs:
  if (start_type == IS_DOUBLE || end_type == IS_DOUBLE) {
    // ... process numeric ranges (floats)
\end{minted}
\Description{Type Confusion in PHP: The vulnerability occurs when the arithmetic condition start\_type + end\_type < 2*IS\_STRING fails due to specific type combinations, causing the program to incorrectly treat non-string data types as strings through Z\_STRVAL\_P() macro calls.}
\caption{Type Confusion Bug in PHP Project}
\label{fig:type_confusion}
\end{figure}

\section{Motivation: \testname Test}\label{sec:motivation}

This section presents a motivating example of a 
minimum viable yet incorrect patch that
would pass the conventional test suite-based patch validation.
We subsequently introduce \testname tests and demonstrate their effectiveness in exposing the flaws of such patches.

\begin{figure}[t]
    \centering
    \begin{subfigure}{\columnwidth}
        \centering
        \begin{minted}[fontsize=\scriptsize,linenos,breaklines, mathescape,numbersep=5pt,xleftmargin=5pt]{diff}
@@ -2924,8 +2924,8 @@ PHP_FUNCTION(range)
 /* If the range is given as strings, 
    generate an array of characters. */
 if (start_type >= IS_STRING || end_type >= IS_STRING) {
-  if (start_type + end_type < 2*IS_STRING) {
+  if (start_type < IS_STRING || end_type < IS_STRING) {
     if (start_type < IS_STRING) {
       if (end_type != IS_ARRAY) {
         php_error_docref(NULL, E_WARNING, "...");
        \end{minted}
        \caption{Developer Patch}
        \label{fig:range_developer_patch}
    \end{subfigure}
    \begin{subfigure}{\columnwidth}
        \centering
        \begin{minted}[fontsize=\scriptsize,linenos,breaklines, mathescape,numbersep=5pt,xleftmargin=5pt]{diff}
@@ -2960,6 +2960,14 @@ PHP_FUNCTION(range)
   }
 
   /* Generate array of characters */
+  if (Z_TYPE_P(user_start) != IS_STRING) {
+    zend_argument_value_error(1, "must be a string");
+    RETURN_THROWS();
+  }
+  if (Z_TYPE_P(user_end) != IS_STRING) {
+    zend_argument_value_error(2, "must be a string");
+    RETURN_THROWS();
+  }

   unsigned char low = Z_STRVAL_P(user_start)[0];
   unsigned char high = Z_STRVAL_P(user_end)[0];
        \end{minted}
        \caption{AVR-Generated Patch}
        \label{fig:range_alternative_patch}
    \end{subfigure}
    \Description{Two distinct patch strategies for the type confusion vulnerability in PHP's \texttt{range()} function.}
    \caption{Two distinct patch strategies for the type confusion vulnerability in PHP's \texttt{range()} function.}
    \label{fig:range_fixes}
\end{figure}

\subsection{Plausible Patch}

We use a type confusion vulnerability (Issue \#13094 \cite{php_issue_13094}) in the PHP interpreter to demonstrate plausible patches. As depicted in \autoref{fig:type_confusion}, this code demonstrates a type confusion vulnerability in PHP's \texttt{\small range()} function that occurs due to flawed type checking logic. The bug arises when the function receives arguments of specific type combinations, such as a double (floating-point number) and an array. The first condition (line 9) correctly identifies that at least one argument appears to be string-like since
$\texttt{IS\_ARRAY}(7) \ge \texttt{IS\_STRING}(6)$,
so the code enters the string-handling branch.
However,
the inner arithmetic condition (line 12) unexpectedly fails when
for example, $\texttt{IS\_DOUBLE}(5)+\texttt{IS\_ARRAY}(7)=12$.
When this condition fails, the code skips the proper type conversion logic and incorrectly attempts to access the non-string data using string accessor macros (line 19), which cause the program crash. 
Two patches exist to address this vulnerability:
one crafted by the developer and
another representing a plausible solution generated by AVR tools.

\noindent\textbf{Developer Patch.} The developer patch (\autoref{fig:range_developer_patch}) addresses the bug by correcting the flawed arithmetic condition,
ensuring that mixed-type inputs are properly redirected to the numeric handling branch instead of 
entering the string processing path.

\noindent\textbf{AVR-Generated Patch.}
The patch (\autoref{fig:range_alternative_patch})
generated by AVR tool
takes a defensive programming approach
by adding explicit type validation at the point of string access.
It inserts runtime checks to verify that
both arguments are actually strings
before attempting to dereference them using \texttt{\small Z\_STRVAL\_P()},
throwing appropriate error messages if non-string types are encountered.

\noindent\textbf{Comparison.} When evaluated against the existing PHP functional test suite and PoC exploits, both patches successfully pass all tests and effectively mitigate the vulnerability.
Automated test suite-based validation would therefore conclude that
both patches are correct.
However,
manual inspection reveals fundamental differences
in their control flow behavior.
The developer patch conditionally
redirects mixed-type inputs to follow the numeric conversion path.
In contrast, the AVR-generated patch immediately terminates execution
with an error for any mixed-type input.
This divergence in control flow indicates that the patches are not functionally equivalent---one may introduce unintended behavioral changes,
which are not covered by any of the existing tests.

\begin{figure}[t]
\begin{minted}[fontsize=\scriptsize,linenos,breaklines,numbersep=5pt,xleftmargin=1pt,escapeinside=||]{php}
|\textbf{---TEST---}|
GH-13094 (range(9.9, '0') causes segmentation fault)
|\textbf{---FILE---}|
<?php
var_dump(range(9.9, '0'));
?>
|\textbf{---EXPECT---}|
|\textbf{\textcolor[HTML]{0B8000}{array}}|(|\textcolor[HTML]{666666}{10}|) {
  [|\textcolor[HTML]{666666}{0}|]=>|\textbf{\textcolor[HTML]{0B8000}{float}}|(|\textcolor[HTML]{666666}{9.9}|)
  [|\textcolor[HTML]{666666}{1}|]=>|\textbf{\textcolor[HTML]{0B8000}{float}}|(|\textcolor[HTML]{666666}{8.9}|)
  [|\textcolor[HTML]{666666}{2}|]=>|\textbf{\textcolor[HTML]{0B8000}{float}}|(|\textcolor[HTML]{666666}{7.9}|)
  [|\textcolor[HTML]{666666}{3}|]=>|\textbf{\textcolor[HTML]{0B8000}{float}}|(|\textcolor[HTML]{666666}{6.9}|)
  [|\textcolor[HTML]{666666}{4}|]=>|\textbf{\textcolor[HTML]{0B8000}{float}}|(|\textcolor[HTML]{666666}{5.9}|)
  [|\textcolor[HTML]{666666}{5}|]=>|\textbf{\textcolor[HTML]{0B8000}{float}}|(|\textcolor[HTML]{666666}{4.9}|)
  [|\textcolor[HTML]{666666}{6}|]=>|\textbf{\textcolor[HTML]{0B8000}{float}}|(|\textcolor[HTML]{666666}{3.9000000000000004}|)
  [|\textcolor[HTML]{666666}{7}|]=>|\textbf{\textcolor[HTML]{0B8000}{float}}|(|\textcolor[HTML]{666666}{2.9000000000000004}|)
  [|\textcolor[HTML]{666666}{8}|]=>|\textbf{\textcolor[HTML]{0B8000}{float}}|(|\textcolor[HTML]{666666}{1.9000000000000004}|)
  [|\textcolor[HTML]{666666}{9}|]=>|\textbf{\textcolor[HTML]{0B8000}{float}}|(|\textcolor[HTML]{666666}{0.9000000000000004}|)
}
\end{minted}
\Description{An example \testname test derived from the PoC for PHP issue \#13094, validating correct behavior for a previously crashing input.}
\caption{An \testname test derived from the PoC for PHP issue \#13094, validating correct behavior for a crashing input.}
\label{fig:poc_derived_test_range}
\end{figure}

\subsection{\testname Tests Reveal the Difference}
 
To find evidence that the AVR-generated patch is incorrect, we observe that the developers not only provided the patch but also created a new test when fixing the bug, as is typical practice \cite{mi2016empirical}, 
and the test case is shown in \autoref{fig:poc_derived_test_range}.
This test case follows the PHP Test (PHPT) format. The format is a structured test specification that consists of several sections: the \texttt{\small --TEST--} section provides a descriptive name for the test case, the \texttt{\small --FILE--} section contains the actual PHP code to be executed, and the \texttt{\small --EXPECT--} section defines the exact output that should be produced. Given that the PHP code within this new test case closely resembles the original PoC, we refer to it as a \testname  test. When we run the \testname  tests on the program with the AVR-generated patch applied, we obtain the following output:
\begin{minted}[fontsize=\scriptsize,breaklines]{text}
Fatal error: Uncaught ValueError: range(): 
Argument #1 ($start) must be a valid string in /test.php:2
Stack trace:
#0 /test.php(2): range(9.9, '0')
#1 {main}
  thrown in /test.php on line 2
\end{minted}
This output differs significantly from the expected output. According to the PHP specification~\cite{php_spec}, the \texttt{\small range()} function is designed to be permissive with respect to argument types: when given a mixture of numeric and string arguments, it performs type coercion and generates a numeric range if either argument is numeric. For example, a call such as \texttt{\small range(9.9, "0")} is valid and expected to produce a descending array of floating-point numbers, as demonstrated in the expected output of the \testname  test case. The developer’s patch preserves this intended behavior by redirecting mixed-type arguments to the numeric range generation logic. In contrast, the AVR-generated patch breaks this specification: by enforcing that both arguments must be strings in the string-handling branch, it rejects mixed-type inputs and raises a runtime error, thereby violating the established PHP semantics. Thus, the AVR-generated patch is incorrect, as it introduces 
a change that deviates from the PHP language specification.
The \testname  test automatically exposes this deviation,
thus demonstrating the value of such tests
for automated patch validation.
%
In other words,
when evaluating an AVR system,
the generated patch should be validated against the
post-patch test suite instead of the pre-patch test suite.

%% file: sections/4_dataset.tex
\begin{table}[t]\footnotesize
\tabcolsep=4pt
\centering
\caption{Overview of Projects and Vulnerabilities in \dataset}
\label{tab:dataset_overview}

\begin{subtable}[t]{\linewidth}
\centering
\begin{tabular}{lccc|lccc}
\toprule
\textbf{Project} & \textbf{LoC} & \textbf{\#} & \textbf{Test} & \textbf{Project} & \textbf{LoC} & \textbf{\#} & \textbf{Test} \\
\midrule
php~\cite{php}            & 1390.2K & 43 & 18.7K  & vim~\cite{vim}            & 564.2K  & 11 & 5.2K  \\
cpython~\cite{cpython}    & 745.9K  & 33 & 48.6K  & hdf5~\cite{hdf5}          & 1334.4K & 8  & 0.6K  \\
llvm~\cite{llvm}          & 8980.4K & 26 & 128.7K & exiv2~\cite{exiv2}        & 93.5K   & 7  & 0.3K  \\
v8~\cite{v8}              & 6225.6K & 24 & 53.7K  & wabt~\cite{wabt}          & 514.9K  & 5  & 1.1K  \\
libxml2~\cite{libxml2}    & 200.4K  & 19 & 3.3K   & hermes~\cite{hermes}      & 590.0K  & 4  & 2.3K  \\
icu~\cite{icu}            & 1241.5K & 15 & 2.0K   & pcap++~\cite{pcapplusplus}& 160.0K  & 3  & 0.3K  \\
quickjs~\cite{quickjs}    & 78.8K   & 2  & 79.7K  & libtiff~\cite{libtiff}    & 109.0K  & 1  & 0.2K  \\
mruby~\cite{mruby}        & 152.4K  & 2  & 1.7K   & jasper~\cite{jasper}      & 5.5K    & 1  & 0.2K  \\
jq~\cite{jq}              & 4.7K    & 2  & 0.9K   & simdjson~\cite{simdjson}  & 547.5K  & 1  & 0.1K  \\
htslib~\cite{htslib}      & 108.3K  & 1  & 0.4K   & wireshark~\cite{wireshark}& 6088.9K & 1  & 0.1K  \\
\midrule
\multicolumn{4}{l}{\textbf{Total Vulnerabilities: 209}} & \multicolumn{4}{l}{} \\
\bottomrule
\end{tabular}
\label{tab:project_overview}
\end{subtable}

\vspace{0.5em}

\begin{subtable}[t]{\linewidth}
\centering
\begin{tabular}{lcc|lcc}
\toprule
\textbf{CWE} & \textbf{\#} & \textbf{Description} & \textbf{CWE} & \textbf{\#} & \textbf{Description} \\
\midrule
CWE-476 & 52 & NULL Dereference  & CWE-670 & 3 & Incorrect Control Flow \\
CWE-617 & 40 & Reachable Assertion       & CWE-415 & 3 & Double Free \\
CWE-122 & 34 & Heap Overflow      & CWE-704 & 3 & Type Confusion \\
CWE-416 & 32 & Use After Free            & CWE-457 & 1 & Uninitialized Memory \\
CWE-190 & 26 & Integer Overflow          & CWE-362 & 1 & Race Condition \\
CWE-121 & 13 & Stack Overflow     & CWE-369 & 1 & Divide by Zero \\
\bottomrule
\end{tabular}
\end{subtable}
\end{table}

\section{\testname Test Dataset \& Validation}

This section presents the \dataset dataset and the validation method of different \testname tests.


\subsection{\dataset Overview}

\noindent\textbf{Vulnerability Statistic.}
\dataset provides a benchmark by incorporating 209 real-world vulnerabilities from 20 open-source projects. These projects represent widely-used systems with extensive codebases and robust test suites, ensuring the vulnerabilities reflect security issues encountered in production environments. As shown in \autoref{tab:dataset_overview}, \dataset covers a diverse range of 12 CWEs, with memory safety issues being most prevalent, including NULL Dereference, UAF, and Heap Overflow, alongside control flow vulnerabilities such as Reachable Assertion and various data handling issues including Integer Overflow.

\noindent\textbf{Selection Criteria.} We identify open-source projects and vulnerabilities for \dataset{} by following a systematic workflow. First, we identified open-source projects based on GitHub star counts, prioritizing those with substantial vulnerability histories that are well-documented through GitHub issues or dedicated vulnerability tracking systems. This process led to the selection of 20 high-quality projects. Subsequently, we conducted a manual review of vulnerabilities reported in these projects over the past ten years, collecting cases that satisfy the following three requirements:

\noindent\textbf{Reproducibility:} Each vulnerability must include build scripts and at least one PoC to enable compilation and reproduction.

\noindent\textbf{Test Coverage:} Each vulnerability must be accompanied by a functional test suite and \testname tests developed by the maintainers.

\noindent\textbf{Functional Preservation:} The vulnerability have been fixed and the set of functionalities remain unchanged before and after applying the fix, ensuring that patches only resolve security vulnerabilities rather than adding a new feature. This requirement is necessary because otherwise \testname tests may validate functionality that does not exist in the unpatched program.

\begin{table}[t]\footnotesize
\tabcolsep=10pt
\centering
\caption{\testname Test Category Distribution by Projects}

\begin{tabular}{ll}
\toprule
\textbf{Category} & \textbf{Projects}  \\
\midrule
     
Output Checking  & exiv2, hermes, htslib, jasper, libxml2, php \\

& jq, llvm-project, simdjson, wabt, wireshark \\
\midrule
Intermed. Checking 
& hdf5, icu, pcapplusplus, libtiff \\
\midrule
Self  Checking         
& cpython, mruby, quickjs, v8, vim \\
\bottomrule
\end{tabular}
\label{tab:poc_test}
\end{table}


\subsection{How Developers Create \testname Tests}

The \testname test is derived from a PoC for the  vulnerability. Unlike common PoCs, which typically only observe if the program crashes, the \testname test performs more comprehensive validations to determine if the program behaviors are as expected. These validations include observing the output content or other intermediate running results. Based on the specific behaviors validated by the \testname, we classified the \testname tests into three categories: \textit{Output Checking}, \textit{Intermediate  Checking} and \textit{Self  Checking}. The distribution of the three categories across the 20 projects is shown in \autoref{tab:poc_test}.

\begin{figure}[t]
    \centering
    \begin{minipage}{\columnwidth}
        \centering
        \begin{minted}[
            fontsize=\scriptsize,
            linenos=false,
            frame=none,
            framesep=2mm,
            bgcolor=gray!10
        ]{python}
class issue_2377_buffer_overflow(metaclass=CaseMeta):
    filename = "$data_path/issue_2377_poc.mp4"
    commands = ["$exiv2 $filename"]
    retval   = [253]
    stderr   = ["$filename: No Exif data found in the file\n"]
    stdout   = ["File name: $filename\nFile size: 225 Bytes\n..."]
        \end{minted}
    \end{minipage}
    \begin{minipage}{\columnwidth}
        \centering
\begin{minted}[
    fontsize=\scriptsize,
    linenos=false,
    frame=none,
    framesep=2mm,
    bgcolor=gray!10,
    escapeinside=||
]{text}
|\textcolor{gray}{// CHECK-LABEL: func.func @nested\_muli() -> i32 \{}|
|\textcolor{gray}{// CHECK:         \%[[VAL\_0:.*]] = "test.constant"() ...}|
|\textcolor{gray}{// CHECK:         \%[[VAL\_1:.*]] = arith.muli \%[[VAL\_0]], ...}|
|\textcolor{blue!70!black}{\textbf{func}}|.|\textcolor{blue!70!black}{\textbf{func}}| |\textcolor{purple}{\textbf{@nested\_muli}}|() -> (|\textcolor{orange!80!black}{\textbf{i32}}|) {
  |\textcolor{red!70!black}{\textbf{\%0}}| = |\textcolor{orange!70!black}{"test.constant"}|() {|\textcolor{blue!70!black}{value}| = |\textcolor{cyan!50!black}{0x7fffffff}|} : () -> |\textcolor{orange!80!black}{\textbf{i32}}|
  |\textcolor{red!70!black}{\textbf{\%1}}| = |\textcolor{orange!70!black}{"test.constant"}|() {|\textcolor{blue!70!black}{value}| = |\textcolor{cyan!50!black}{-2147483648}|} : () -> |\textcolor{orange!80!black}{\textbf{i32}}|
  ...
}
\end{minted}
    \end{minipage}
    \begin{minipage}{\columnwidth}
        \centering
        \begin{minted}[
            fontsize=\scriptsize,
            linenos=false,
            frame=none,
            framesep=2mm,
            bgcolor=gray!10
        ]{xml}
<!-- oss-fuzz-51295_0.xsd (input) -->
<xs:schema xmlns:xs="http://www.w3.org/2001/XMLSchema">
    <xs:element name="e" substitutionGroup="e"/>
</xs:schema>

<!-- oss-fuzz-51295_0.err (expected error output) -->
element decl. 'e': The element declaration 'e' defines 
a circular substitution group to element declaration 'e'.
        \end{minted}
    \end{minipage}
    
    \Description{\testname Examples for Output Checking: Examples from Exiv2 (top), LLVM (middle), and libxml2 (bottom) showing how tests specify input and validate expected output.}
    \caption{\testname Test Examples for Output Checking}
    \label{fig:output_test_examples}
\end{figure}

\noindent\textbf{Output Checking.} This category applies to programs that process external input files or byte streams and produce observable output. During patch validation, the \testname test executes the program with the AVR-generated patch applied, feeds it the original PoC input, and compares the actual output against the expected output stored as part of the test. The structure of \textit{Output Checking} varies by project. As illustrated in the motivation example (\autoref{fig:poc_derived_test_range}), PHP's \testname tests use the PHPT format. \autoref{fig:output_test_examples} demonstrates three other representative formats: Exiv2 employs a Python-based framework specifying the input file, command, expected return value, and stderr/stdout content; LLVM uses inline CHECK directives to verify generated intermediate representation; and libxml2 separates input XML schemas from expected error messages in distinct files. Despite these differences, all formats share a common structure of defining inputs and expected outputs.

The testcase production process for this category is straightforward: compile the program with the correct patch applied, execute it with the PoC as input to capture the expected output, and format the result according to the project's test framework conventions.
We observed that several projects in \dataset,
including Hermes, libxml2, LLVM, PHP, and Wabt,
have implemented automated scripts using similar methodologies.
The locations of these production scripts are provided in \autoref{tab:auto_gen}.

\noindent\textbf{Intermediate  Checking.}  This category applies to vulnerabilities in library codebases where the original PoC is a source file (i.e., a harness) that invokes a sequence of API functions. During the patch validation stage, the \testname test executes a modified harness that encodes expected intermediate results and determines success based on the return value. Consider the $\text{PoC} \rightarrow~$\testname transformation for a bug in the HDF5 library, as illustrated in \autoref{fig:intermediate_poc_plus}. The original PoC consists of a sequence of API calls. The \testname test modifies this harness by inserting $\texttt{CHECK}$ and $\texttt{VERIFY}$ macros to assert expected intermediate behavior at each step. Specifically, the bug-triggering call is checked by the assertion \texttt{VERIFY(ret, FAIL, ...)}. This structure ensures that the patched function correctly detects the invalid input state and returns the appropriate error code ($\texttt{FAIL}$) rather than crashing, thereby validating the patch.

Developers produce these test cases by instrumenting API calls to capture both return values and pointer-based outputs at runtime. Since C lacks multiple return values, APIs often use pointer arguments for additional outputs. Checking logic is then inserted at an appropriate abstraction level to validate these captured values.

\begin{figure}[t]
    \centering
    \begin{minipage}{0.48\textwidth}
        \centering
        \begin{minted}[
            fontsize=\scriptsize,
            linenos=false,
            frame=none, % Remove all frames/borders
            framesep=2mm,
            bgcolor=gray!10
        ]{c}
// Original PoC for hdf5 bug
space_id = H5Screate_simple(1, dims, NULL);
H5Sselect_hyperslab(space_id, ...);
H5Sset_extent_none(space_id);
H5Sget_select_hyper_blocklist(space_id, ...); // Vulnerability Triggered Here
H5Sclose(space_id);
        \end{minted}
    \end{minipage}
    \hfill
    \begin{minipage}{0.48\textwidth}
        \centering
        \begin{minted}[
            fontsize=\scriptsize,
            linenos=false,
            frame=none,
            framesep=2mm,
            bgcolor=green!10
        ]{c}
static void poc_plus_test(void) {
  hsize_t dims[]  = {10}; /* ... initialization ... */
  space_id = H5Screate_simple(1, dims, NULL);
  CHECK(space_id, H5I_INVALID_HID, "H5Screate_simple");
  ret = H5Sselect_hyperslab(space_id, ...);
  CHECK(ret, FAIL, "H5Sselect_hyperslab");
  ret = H5Sset_extent_none(space_id);
  CHECK(ret, FAIL, "H5Sset_extent_none");
  ret = H5Sget_select_hyper_blocklist(space_id, ...);
  VERIFY(ret, FAIL, "H5Sget_select_hyper_blocklist");
  ret = H5Sclose(space_id);
  CHECK(ret, FAIL, "H5Sclose");
}
        \end{minted}
    \end{minipage}
    \Description{\testname Transformation for Intermediate  Checking: The process involves inserting CHECK and VERIFY macros into the harness to ensure intermediate API calls return the expected results after the patch is applied.}
    \caption{\testname Test Example for Intermediate  Checking}
    \label{fig:intermediate_poc_plus}
\end{figure}

\noindent\textbf{Self Checking.}
This category targets interpreter programs, such as Python, JavaScript engines, or Ruby interpreters, where vulnerabilities are triggered by executing code written in the interpreted language. The goal of \testname construction is to transform the original PoC script into a self-validating test that explicitly verifies expected runtime behavior. Rather than merely observing whether the interpreter crashes, the \testname test embeds assertions within the interpreted program itself to confirm that the patched interpreter properly handles previously vulnerable inputs by raising appropriate exceptions, producing specific error messages, or returning expected values. This transformation requires understanding the correct post-patch behavior and inserting corresponding exception handlers and assertions to validate the interpreter's response. 
Consider the example transformation for a CPython bug illustrated in \autoref{fig:self_test_example}. The original PoC consists of simple function calls that trigger a crash in the unpatched interpreter. The \testname test transforms this into a self-validating script by wrapping each vulnerable call in a \texttt{self.assertRaises(TypeError)} context manager, verifying both that the expected exception type is raised. This structure ensures that the patched interpreter correctly detects malformed input and raises appropriately typed exceptions.

To produce self-checking tests, developers first execute the original PoC on the patched interpreter to observe the corrected behavior, whether it raises a specific exception, returns a particular value, or outputs an error message. They may also transform the PoC to capture all possible error paths, ensuring comprehensive coverage of the fix. Developers then wrap vulnerable code paths in appropriate assertion constructs.
Since the interpreters in \dataset are all about dynamic languages,
developers may also verify side effects---such as changes to global state, object attributes, or resource handles---to confirm that the interpreter maintains correct behavior beyond just the immediate return value.

\begin{figure}[t]
    \centering
    \begin{minipage}{0.48\textwidth}
        \centering
        \begin{minted}[
            fontsize=\scriptsize,
            linenos=false,
            frame=none,
            framesep=2mm,
            bgcolor=gray!10
        ]{python}
# Original PoC for CPython sys bug
import sys
sys.remote_exec(0, None)
        \end{minted}
    \end{minipage}
    \hfill
    \begin{minipage}{0.48\textwidth}
        \centering
        \begin{minted}[
            fontsize=\scriptsize,
            linenos=false,
            frame=none,
            framesep=2mm,
            bgcolor=green!10
        ]{python}
# PoC+ test with self-test
import sys
with self.assertRaises(TypeError):
    sys.remote_exec(0, None)
with self.assertRaises(TypeError):
    sys.remote_exec(0, 123)
\end{minted}
    \end{minipage}
    \Description{\testname Transformation for Self  Checking: The process involves wrapping vulnerable function calls with exception handlers and adding assertions to verify that the patched interpreter raises the expected exception types with appropriate error messages.}
    \caption{\testname Test Example for Self Checking}
    \label{fig:self_test_example}
\end{figure}

%% file: sections/5_overestimation.tex



\section{Quantifying Overestimation}\label{sec:overestimation}

In this section, we examine the extent to which conventional test suite-based validation overestimates AVR system effectiveness.

\subsection{Methodology}
We evaluated three state-of-the-art AVR systems: \patchagent \cite{PatchAgent}, \santopatch \cite{SAN2PATCH}, and SWE-Agent\cite{yang2024sweagent} on \dataset to quantify the extent to which test suite-based validation overestimate AVR system performance. All tools use the LLM-based approach. Since the original SWE-Agent does not support C/C++ programs, we use its multi-language version \cite{zan2025multi}. For each tool, we conducted experiments with two different large language models: GPT-4.1 version \textit{gpt-4.1-2025-04-14} \cite{openai} and Claude-4 Sonnet version \textit{claude-sonnet-4-20250514} \cite{anthropic}. To  account for the non-deterministic nature of LLM-based patch generation, we executed each configuration,
i.e., a $($AVR tool, LLM version$)$ pair,
five times on every test case,
resulting in 1045 (209 $\times$ 5) patch attempts
per each configuration for all 209 vulnerabilities,
%
where each vulnerability will receive at most 30
(6 configurations $\times$ 5 attempts) patches.

Our experiment employs a two-stage validation framework
designed to expose the limitations of conventional validation approaches.
In Stage 1 (Basic Validation),
generated patches are validated using the conventional approach:
verifying PoC mitigation and
executing the project's existing functional test suite.
Patches passing both criteria are classified as ``correct''
under conventional validation.
In Stage 2 (\testname Validation),
patches deemed correct in Stage 1 are further evaluated
using \testname tests to assess
whether the patch also conforms to
the semantics or developers' intention
encoded in the \testname test.
This stage reveals false positives,
i.e., patches that appear correct under basic tests
but fail on \testname tests. 

\subsection{General Results}
\label{subsec:general-results}

Our experiment results reveal a substantial gap
between conventional test suite-based validation and
rigorous \testname test validation.
As shown in \autoref{tab:general_results},
\patchagent with GPT-4.1 generated 798 patches that passed basic tests
among 1045 executions (209 vulnerabilities $\times$ 5 attempts),
achieving an initial success rate of 76.4\%.
Similarly, \patchagent with Claude Sonnet-4
achieved an initial success rate of 83.5\%
%
However,
when these seemingly correct patches were subjected to \testname test validation, the success rates dropped dramatically to 44.5\% and 50.1\%, respectively,
exposing false discovery rates (FDR) of 41.7\% and 40.1\%.

\santopatch demonstrated lower overall patch generation rates
but similar validation reliability issues,
with initial success rates of
37.9\% (GPT-4.1) and 41.3\% (Sonnet-4)
declining to 19.6\% and 20.7\% under \testname validation,
resulting in FDRs of 48.2\% and 49.8\%.
SWE-Agent exhibited the poorest performance
with initial success rates of only 14.4\% (GPT-4.1) and 29.0\% (Sonnet-4), further declining to 8.3\% and 19.6\% respectively
under \testname validation. 
The full results are provided in \autoref{fig:full_results}.

\begin{figure}[t]
    \centering
    \begin{subfigure}{\columnwidth}
        \centering
        \includegraphics[width=\textwidth]{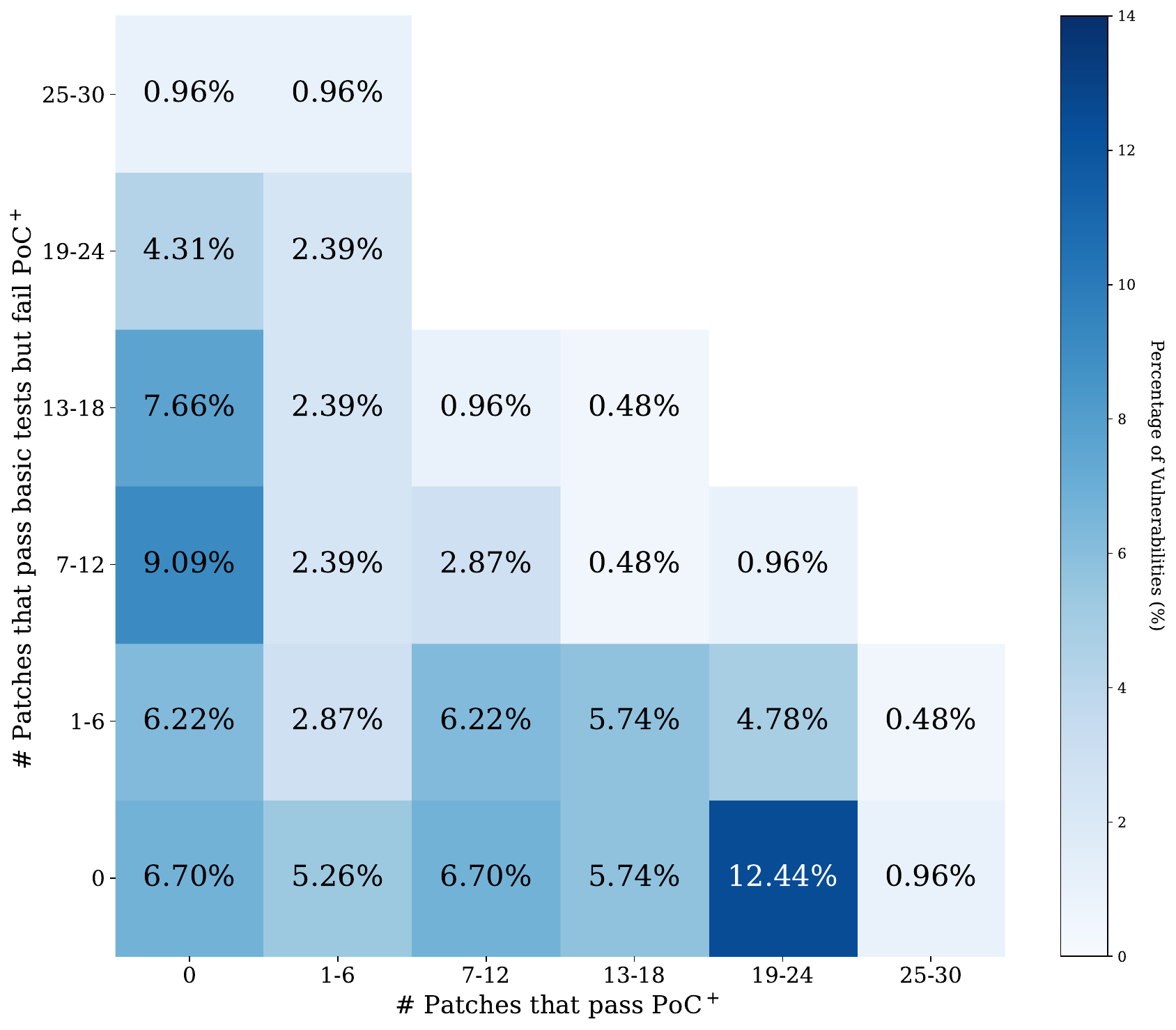}
        \label{fig:vuln_heatmap}
    \end{subfigure}
    \begin{subfigure}{\columnwidth}
        \centering
        \begin{minipage}{0.9\columnwidth}
            \centering
            \includegraphics[width=\textwidth]{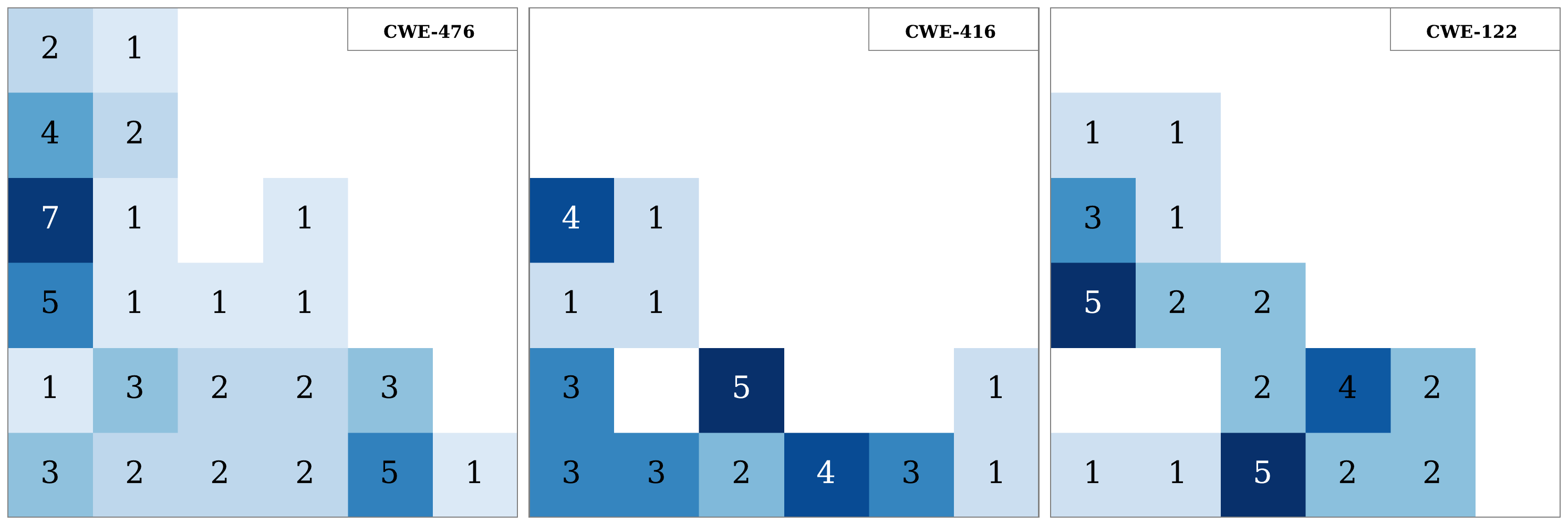}
        \end{minipage}
        \hfill
        \begin{minipage}{0.9\columnwidth}
            \centering
            \includegraphics[width=\textwidth]{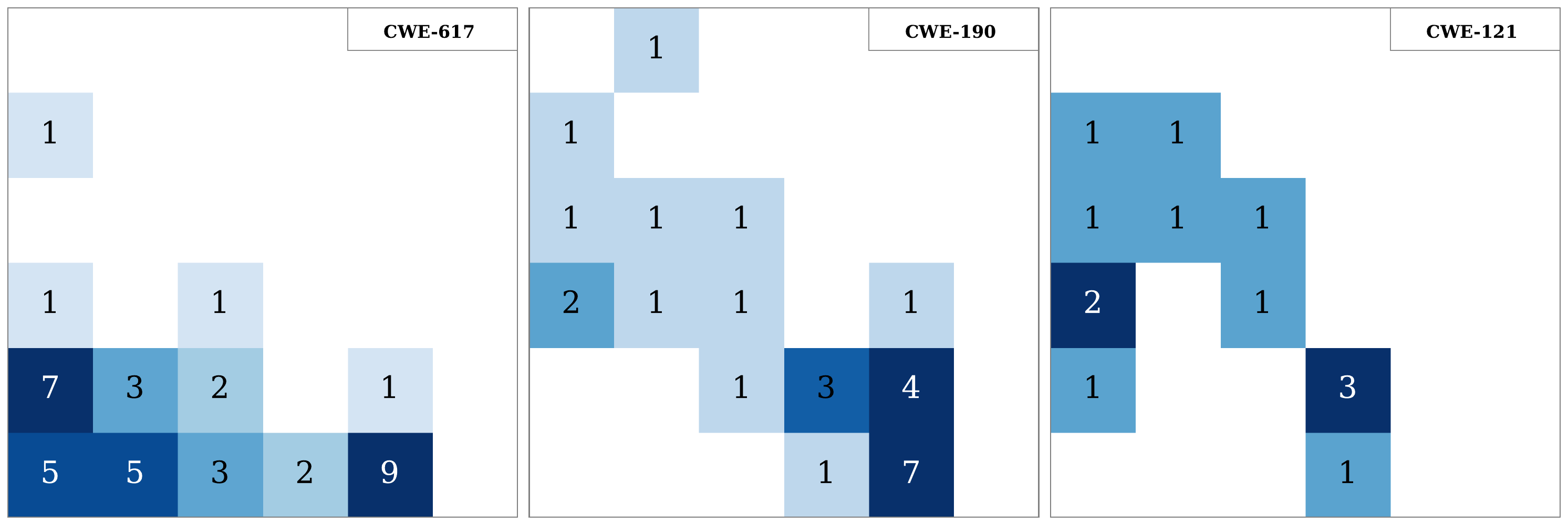}
            \label{fig:second_graph}
        \end{minipage}
    \end{subfigure}
    \Description{Distribution of vulnerabilities: The x-axis shows number of patches passing \testname tests, while the y-axis shows number of patches passing basic tests but failing \testname tests.}
    \caption{Distribution of vulnerabilities: The x-axis shows \#patches passing \testname tests, while the y-axis shows \#patches passing basic tests but failing \testname tests.}
    \label{fig:vulnerability_matrix}
\end{figure}

Across all configurations,
our evaluation consistently reveals an FDR around 40\%,
i.e., about 40\% of patches that pass basic tests fail on \testname tests.
To analyze the repair outcomes for different vulnerabilities in our experiments. \autoref{fig:vulnerability_matrix} illustrates the distribution of vulnerabilities based on the ratio of correct patches to false positive patches generated, revealing four distinct behavioral patterns:

\begin{table}[t]\small
    \centering
    \caption{Performance of AVR tools under different validation. \textbf{init}: patches passing basic tests; \textbf{poc+}: patches also passing \testname tests; 
    \textbf{FDR}: false discovery rate, i.e., the fraction of initially validated patches that fail subsequent testing.}

    \begin{tabular}{l|c|cccc}
        \toprule
        \textbf{Tool} & \textbf{Model} & \textbf{init} & 
        \textbf{poc+}  & \textbf{FDR} \\
        \midrule
        \multirow{2}{*}{\patchagent} 
                      & Sonnet-4 & 83.5\%     
                      & 50.1\% & 40.1\% (350/873) \\
                      & GPT-4.1  & 76.4\%     
                      & 44.5\% & 41.7\% (333/798) \\
        \midrule
        \multirow{2}{*}{\santopatch}
                      & Sonnet-4 & 41.3\%     
                      & 20.7\% & 49.8\% (215/432) \\
                      & GPT-4.1  & 37.9\%     
                      & 19.6\% & 48.2\% (191/396) \\
        \midrule
        \multirow{2}{*}{SWE-Agent}
                      & Sonnet-4 & 29.0\%     
                      & 19.6\% & 32.3\% (98/303) \\
                      & GPT-4.1  & 14.4\%     
                      & 8.3\%  & 41.3\% (63/150) \\
        \midrule
        \multicolumn{2}{c|}{\textbf{Overall}} & 47.1\%     
                      & 27.1\% & 42.3\% (1250/2952) \\
        \bottomrule
    \end{tabular}
    \label{tab:general_results}
\end{table}

\noindent\textbf{Repair-Resistant Vulnerabilities.} 6.70\% of vulnerabilities cluster at the origin (no patch passes any tests), representing cases where AVR tools completely fail to generate any patches that pass basic validation. These vulnerabilities consistently challenge all tested AVR configurations, indicating limitations in current tools.
    
\noindent\textbf{High-Success Repair Targets.} 31.10\% of vulnerabilities demonstrate excellent repairability, generating correct patches without producing any FPs. The distribution peaks at 12.44\%, representing repair scenarios with 19-24 correct patches and zero FPs, while an additional 6.70\% produce 7-12 correct patches with perfect accuracy. 
    
\noindent\textbf{False-Positive Prone Vulnerabilities:} 28.24\% of vulnerabilities generate exclusively false positive patches while producing zero genuinely correct patches that pass comprehensive testing. The distribution peaks at 9.09\% for cases producing 7-12 false positives with no correct patches. These scenarios are particularly concerning for real-world deployment, as AVR tools consistently generate plausible but fundamentally flawed fixes, creating a dangerous illusion of successful repair that could introduce new security risks.
    
\noindent\textbf{Mixed-Performance Vulnerabilities:} 33.96\% of vulnerabilities exhibit mixed repair patterns across intermediate performance regions, generating varying combinations of correct and incorrect patches in low-to-moderate quantities. 

The vulnerability distribution reveals a complex, multi-modal pattern
demonstrating varied AVR tool performance across real-world vulnerabilities.
Rather than showing a continuous distribution,
the data exhibits distinct clustering with 6.7\% producing no patches,
31.10\% achieving strong repair outcomes, and
28.24\% generating only false positives.
%
This clustering pattern suggests that
repairability is predominantly
an intrinsic characteristic of the vulnerability itself rather than
a consequence of specific tool configurations or parameter choices.

\begin{table*}[t]
\centering
\caption{Manual Categorization of Patches that Passed \testname Tests Across Different AVR Tools: The table shows the distribution of four quality categories
Results are compared across three tools 
using two LLMs 
.}

\tabcolsep=7pt
\begin{tabular}{lccccccc}
\toprule
\multirow{2}{*}{\textbf{Category}} & \multicolumn{2}{c}{\textbf{\patchagent}} & \multicolumn{2}{c}{\textbf{\santopatch}} & \multicolumn{2}{c}{\textbf{SWE-Agent}} & \multirow{2}{*}{\textbf{Total}} \\
\cmidrule(lr){2-3} \cmidrule(lr){4-5} \cmidrule(lr){6-7}
& \textbf{Sonnet-4} & \textbf{GPT-4.1} & \textbf{Sonnet-4} & \textbf{GPT-4.1} & \textbf{Sonnet-4} & \textbf{GPT-4.1} & \\
\midrule
Semantic Equivalent  & 396 (75.72\%) & 339 (72.90\%) & 156 (71.89\%)  & 150 (73.17\%) & 159 (77.56\%) & 66 (75.86\%) & 1266 (74.38\%) \\
Performance Issue    &  25  (4.78\%) & 24 (5.16\%)   &   1 (0.46\%)   &   3 (1.46\%)  & 3 (1.46\%)    & 0 (0.00\%) & 56 (3.29\%) \\
Suboptimal Repair    &  59 (11.28\%) & 61 (13.12\%)  &  28 (12.90\%)  &  27 (13.17\%) & 20 (9.76\%)  & 13 (14.94\%) & 208 (12.22\%) \\
Check Circumvention  &  43  (8.22\%) & 41 (8.82\%)   &  32 (14.75\%)  &  25 (12.20\%) & 23 (11.22\%) & 8 (9.20\%) & 172 (10.11\%) \\
\midrule
\textbf{Total} & \textbf{523 (100\%)} & \textbf{465 (100\%)} & \textbf{217 (100\%)} & \textbf{205 (100\%)} & \textbf{205 (100\%)} & \textbf{87 (100\%)} & \textbf{1702 (100\%)} \\
\bottomrule
\end{tabular}
\label{tab:true_positive_category}
\end{table*}

To understand these clustering patterns by CWE category,
we conducted detailed analysis on
six prevalent ($\ge$ 10 samples) CWE types in our dataset.
CWE-617 (Reachable Assertion) emerges as the most resistant vulnerability type to automated repair, likely because LLMs lack extensive training data on developer-introduced assertions compared to common memory errors. However, when patches are successfully generated for CWE-617, they demonstrate higher reliability than other vulnerability types.
Among memory errors, NULL pointer dereference (CWE-476) and stack-based buffer overflow (CWE-121) both exhibit high false discovery rates. This may occur because these vulnerabilities typically have standardized repair methods, leading LLMs to generate superficial fixes—null checks for CWE-476 and boundary checks for CWE-121—rather than addressing the underlying architectural issues.
Use After Free (CWE-416) and Heap-based Buffer Overflow (CWE-122) demonstrate more reliable repair outcomes. Integer overflow (CWE-190) also employs general fix methods but shows high reliability, possibly because upgrading to larger data types is a common practice in real-world development.

%% file: sections/6_reliability.tex
\section{Reliability of \testname Test}\label{sec:true_positive}

In this section, we examine the reliability of \testname tests as a validation method for evaluating patch correctness.

\subsection{Methodology}
To assess the reliability of \testname tests and identify potential issue 
of using them for AVR tools evaluation,
we conducted a manual review of all patches that passed \testname tests by
comparing them with developer patches.
The results are presented in \autoref{tab:true_positive_category}.
We classified these patches into four categories
based on their 
semantic closeness
to the developer patches.
Two patches are considered semantically equivalent
if they satisfy the following criteria:

\begin{itemize}[noitemsep,nolistsep]
\item Both patches target the same function
\item Both patches have identical time and space complexity
\item The logic implemented by both patches is functionally equivalent with no unintended side effects
\end{itemize}


For patches that do not achieve semantic equivalence, we proceed with further categorization. If we find that a patch uses ad-hoc methods to bypass security or functional checks, we classify it as an \emph{Check Circumvention}, as such approaches are in fact 
incorrect implementations despite 
passing all tests, including \testname.
For patches with correct functionality but different performance characteristics, we evaluate whether the AVR-generated solution exhibits worse time or space complexity compared to the developer solution. Such cases are classified as having \emph{performance issues}.
Finally, for patches that are functionally correct and have same algorithm complexity with developers patch, we assess quality by examining whether the developer's patch is more obvious for correctness compared to the AVR-generated patch. 
We classify the AVR-generated patch as having \emph{Suboptimal Repairs}.

We implemented an inter-rater reliability process \cite{appatch,yang2023enhancing}
where each author independently categorized the patches,
and then cross-validated results and discussed any discrepancies
until reaching consensus.
\textbf{Our findings show that more than 70\% of the patches maintain semantic equivalence with developer patches},
demonstrating the high reliability of the \testname tests.
The subsequent sections examine the remaining patch categories in detail.

\begin{figure}[t]
    \centering
    \begin{minted}[fontsize=\tinyfootnote,linenos,breaklines, mathescape,numbersep=5pt,xleftmargin=5pt]{c}
static int template_clear(TemplateObject *self) {
    Py_CLEAR(self->literal);
    for (Py_ssize_t i = 0, n = Py_SIZE(self); i < n; i++)
        // BUG: items[i].literal may be uninitialized
        Py_CLEAR(self->items[i].literal);
    return 0;
}
static PyObject *sre_template_impl(..., PyObject *template) {
    // template is a list containing interleaved 
    // literal strings (str or bytes) and group indices (int)
    // [literal1, group1, literal2, ..., literalN].
    Py_ssize_t n = /* Extract number of groups */;
    // Allocate array but items[].literal not initialized
    TemplateObject *self = PyObject_GC_NewVar(TemplateObject, ..., n);
    self->literal = Py_NewRef(PyList_GET_ITEM(template, 0));
    // FIX OPTION 1: Initialize all literal fields to NULL
    // This prevents crashes since Py_CLEAR safely handles NULL
    // for (Py_ssize_t i = 0; i < n; i++)
    //     self->items[i].literal = NULL;
    for (Py_ssize_t i = 0; i < n; i++) {
        Py_ssize_t index = // Extract (i+1)-th group number;
        if (index < 0) {
            // FIX OPTION 2: Set object size to track 
            // how many items were successfully initialized.
            // template_clear will only clean initialized items.
            // Py_SET_SIZE(self, i);
            goto bad_template;
        }
        // Normal case: would initialize items[i].literal here...
    }
    return (PyObject*) self;
bad_template:
    PyErr_SetString(PyExc_TypeError, "invalid template");
    Py_XDECREF(self); // Triggers template_clear() cleanup
    return NULL;
}
\end{minted}
\Description{Performance Issue Example: Uninitialized Memory Access Bug: The cleanup function template\_clear assumes all items[].literal fields are valid object pointers, but early error conditions can leave them uninitialized.}
\caption{Example of Performance Issue}
\label{fig:uninit_access}
\end{figure}

\subsection{Performance Issue}

Performance issues arise when patches are functionally correct but employ repair strategies with suboptimal algorithmic complexity compared to developer solutions. To evaluate performance differences, we analyze the complexity of the extra operations required specifically for the repair, rather than comparing the overall program complexity. This approach isolates the computational overhead introduced by different repair strategies. If we find that the time or space complexity of extra operations for repairing the bug by the AVR tools is higher than the developer's, we categorize the patch as having a performance issue. It is important to note that categorizing a patch as having a performance issue does not necessarily indicate that it is better or worse than the developer's solution, as the project may not be particularly sensitive to performance considerations. This category simply identifies the difference in computational efficiency between repair approaches.

To illustrate this category, we examine a case from CPython involving an uninitialized memory access vulnerability. \autoref{fig:uninit_access} demonstrates how two algorithmically distinct repair strategies can address the same bug with  different time complexity for their repair operations. The issue occurs when \texttt{\small PyObject\_GC\_NewVar()} allocates memory for a \texttt{\small TemplateObject} but leaves the \texttt{\small items[i].literal} fields uninitialized. If an error condition arises during the initialization loop (such as encountering a negative index), the code jumps to the error handler, which calls \texttt{\small Py\_XDECREF(self)} and subsequently triggers \texttt{template\_clear()}. This cleanup function blindly iterates through all allocated items and calls \texttt{\small Py\_CLEAR()} on potentially uninitialized \texttt{\small literal} fields, causing crashes when the macro attempts to decrement reference counts on garbage memory values.

Two distinct repair strategies emerge to address this vulnerability, which differ significantly in their algorithmic complexity. The AVR-suggested fix takes a defensive programming approach by proactively initializing all \texttt{\small literal} fields to \texttt{\small NULL} immediately after allocation. This repair strategy has $O(n)$ complexity. In contrast, the developer patch employs a tracking-based strategy using \texttt{\small Py\_SET\_SIZE(self, i)} to record how many items have been successfully initialized. This repair approach has $O(1)$ complexity.

\subsection{Suboptimal Repair}\label{sec:sub}
Suboptimal repair represents patches
that are functionally correct and maintain the same algorithmic complexity
as developer solutions,
but exhibit inferior implementation quality that
makes their correctness less apparent.
With no intuitive justification on why
code changes are made at specific locations,
this type of patch eventually hurts the maintainability of the overall codebase.
Patches in this category typically fall into two patterns.

First,
AVR-generated patches often address vulnerabilities
at later stages in the execution flow
rather than preventing the underlying issue at its source.
Similar to the incorrect root cause example
in our motivating case (\autoref{sec:motivation}),
developer patches typically fix vulnerabilities at creation sites
where corrupted data structures are initially formed,
while AVR tools apply defensive measures at usage sites where problems manifest.
The prevention-oriented approach
is more intuitively correct
because it detects problems at their source and
makes the code more self-documenting.

\begin{figure}[t]
\centering
\begin{subfigure}[t]{0.48\textwidth}
\centering
\begin{minted}[fontsize=\tinyfootnote,linenos,breaklines,numbersep=3pt,xleftmargin=1pt]{diff}
@@ -729,10 +729,14 @@ void add_class_vars(...)
     }
     prop = NULL;
     if (statics && (info->flags & ACC_STATIC) != 0) {
-        prop = &ce->static_members_table[info->offset];
+        if (ce->static_members_table && 
+            info->offset < ce->static_members_count) {
+            prop = &ce->static_members_table[info->offset];
+        }
     } else if (!statics && (info->flag & ACC_STATIC) == 0) {
-        prop = &property_table[info->offset];
+        if (property_table && info->offset < ce->property_count) {
+            prop = &property_table[info->offset];
+        }
       }
       if (!prop) {
          continue;
\end{minted}
\caption{AVR-Generated Patch}
\end{subfigure}
\hfill
\begin{subfigure}[t]{0.48\textwidth}
\centering
\begin{minted}[fontsize=\tinyfootnote,linenos,breaklines,numbersep=3pt,xleftmargin=1pt]{diff}
@@ -724,7 +724,8 @@ void add_class_vars(...)
     if (((info->flags & ACC_PROTECTED) &&
          !check_protected(info->ce, scope)) ||
          ((info->flags & ACC_PRIVATE) &&
-           info->ce != scope)) {
+           info->ce != scope) ||
+          (info->flags & ACC_VIRTUAL)) {
           continue;
     }
     prop = NULL;
\end{minted}
\caption{Developer Patch}
\end{subfigure}
\Description{Suboptimal Repair Example: The AVR patch adds bounds checking, while the developer patch demonstrates semantic understanding by excluding virtual properties.}
\caption{Example of Suboptimal Repair}
\label{fig:sub_optimal}
\end{figure}

Second,
developer solutions often encode richer semantic information and domain-specific knowledge
compared to AVR-generated alternatives.
To illustrate this pattern,
consider a heap out-of-bounds vulnerability in PHP's class variable handling
shown in \autoref{fig:sub_optimal}.
The PoC of the vulnerability defines a class with virtual members,
but the handling function incorrectly forgets to handle this case,
causing heap out-of-bounds access.
The AVR-generated patch applies defensive bounds checking
by validating array pointers and offsets before access,
essentially forcing a fix through defensive programming.
In contrast,
developer's patch adds a single condition \texttt{\small (info->flags \& ACC\_VIRTUAL)} to exclude virtual properties from processing,
demonstrating semantic understanding that
virtual properties should not be handled in this context.
While both patches prevent the crash,
developer's solution encodes the actual business logic---virtual properties are conceptually different and require separate handling---making it obviously better despite being difficult to distinguish through automated testing.
This semantic richness reflects deep understanding of PHP's object model and
makes the code more maintainable.

\subsection{Check Circumvention}

Check circumvention represent patches that attempt to bypass security or functional checks rather than addressing the underlying root cause of
bugs---another way of suppressing an error! These patches prioritize immediate test passage over principled problem resolution, often employing workarounds that circumvent protective mechanisms or safety validations. Such approaches fundamentally misunderstand the purpose of security checks and program invariants, treating them as obstacles to avoid rather than indicators of deeper logical issues.
Our analysis identified two primary patterns of check circumvention. First, for out-of-bounds access vulnerabilities, AVR patches frequently resort to excessive memory over-allocation, attempting to prevent crashes by allocating significantly larger buffer sizes rather than determining and allocating the exact memory size the program requires or correcting the faulty indexing logic. Second, for reachable assertion failures, patches commonly either remove assertion statements entirely or artificially manipulate variables immediately before assertions to force conditions to evaluate as true, effectively disabling safety checks without understanding why the assertions were violated.

%% file: sections/7_false_positive.tex
\begin{table*}[htbp]
\centering
\caption{Categorization of FP Patches: The table presents a breakdown of FP patches across three AVR systems using two LLMs.}
\tabcolsep=7.5pt
\begin{tabular}{lccccccc}
\toprule
\multirow{2}{*}{\textbf{Category}} & \multicolumn{2}{c}{\textbf{\patchagent}} & \multicolumn{2}{c}{\textbf{\santopatch}} & \multicolumn{2}{c}{\textbf{SWE-Agent}} & \multirow{2}{*}{\textbf{Total}} \\
\cmidrule(lr){2-3} \cmidrule(lr){4-5} \cmidrule(lr){6-7}
& \textbf{Sonnet-4} & \textbf{GPT-4.1} & \textbf{Sonnet-4} & \textbf{GPT-4.1} & \textbf{Sonnet-4} & \textbf{GPT-4.1} & \\
\midrule
Incorrect Root Cause    & 144 (41.14\%) & 118 (35.44\%) & 113 (52.56\%) & 81 (42.41\%) & 37 (37.76\%) & 22 (34.92\%) & 515 (41.18\%) \\
Specification Violation & 189 (54.00\%) & 200 (60.06\%) & 93 (43.26\%) & 97 (50.79\%)  & 61 (62.24\%) & 40 (63.49\%) & 680 (54.38\%) \\
Poor Code Practice      &  17 (4.86\%)  & 15  (4.50\%)  & 9 (4.19\%) & 13 (6.81\%)     & 0 (0.00\%) & 1 (1.59\%) & 55 (4.40\%) \\
\midrule
\textbf{Total} & \textbf{350 (100\%)} & \textbf{333 (100\%)} & \textbf{215 (100\%)} & \textbf{191 (100\%)} & \textbf{98  (100\%)} & \textbf{63  (100\%)} & \textbf{1250 (100\%)} \\
\bottomrule
\end{tabular}

\label{tab:false_positive_category}
\end{table*}

\section{False Positive Analysis}\label{sec:finding}

We conducted a systematic analysis of all false positives
from our \dataset experiments (\autoref{subsec:general-results})
to understand the reasons why these patches pass the basic tests
but fails \testname.
%
In particular,
we manually compared those patches with developer patches
%
%
and classified them into three categories:
patches that incorrectly identify the root cause of the issue,
patches that violate project specifications, and
patches that use poor coding practices.
%
\autoref{tab:false_positive_category} presents the distribution of these categories across all tested configurations.
Our analysis reveals that \textit{Specification Violation} represents the most prevalent failure mode, accounting for approximately 55.57\% of all false positive cases. This is followed by \textit{Incorrect Root Cause} at 39.88\%, while \textit{Poor Code Practice} constitute the smallest category.

\begin{figure}[t]
\centering
\begin{subfigure}[t]{0.48\textwidth}
\centering
\begin{minted}[fontsize=\scriptsize,linenos,breaklines,numbersep=3pt,xleftmargin=1pt]{diff}
@@ -2234,6 +2234,10 @@ PySequence_Count(...) {
 PySequence_Contains(PyObject *seq, PyObject *ob)
 {
+    if (seq == NULL) {
+        null_error();
+        return -1;
+    }
     PySeqMethods *sqm = Py_TYPE(seq)->tp_as_sequence;
     if (sqm != NULL && sqm->sq_contains != NULL) {
\end{minted}
\caption{AVR-Generated Patch}
\label{fig:llm_patch_wrong_root}
\end{subfigure}
\hfill
\begin{subfigure}[t]{0.48\textwidth}
\centering
\begin{minted}[fontsize=\scriptsize,linenos,breaklines,numbersep=3pt,xleftmargin=1pt]{diff}
@@ -5083,19 +5083,17 @@ ast_type_init(...) {
     PyObject *key, *value, *fields, *attributes = NULL;
-    if (PyObj_GetOptionalAttr((PyObject*)Py_TYPE(self), 
-                        state->_fields, &fields) < 0) {
+    fields = PyObj_GetAttr((PyObject*)Py_TYPE(self), 
+                             state->_fields);
+    if (fields == NULL)
         goto cleanup;
-    if (fields) {
-        numfields = PySequence_Size(fields);
-        if (numfields == -1)
-            goto cleanup;
-        remaining_fields = PySet_New(fields);
-    }
-    else
-        remaining_fields = PySet_New(NULL);
+    numfields = PySequence_Size(fields);
+    if (numfields == -1)
+        goto cleanup;
+    remaining_fields = PySet_New(fields);
     if (remaining_fields == NULL)
         goto cleanup;
\end{minted}
\caption{Developer Patch}
\label{fig:human_patch_correct_root}
\end{subfigure}
\Description{Incorrect Root Cause Example: Comparison of AVR and developer patches for a NULL pointer dereference in CPython's AST module.}
\caption{Incorrect Root Cause Example}
\label{fig:incorrect_root_cause_patches}
\end{figure}

\subsection{Incorrect Root Cause}

We classify patches into this category
when the generated patch modifies code
in a different function from the developer patch,
indicating fundamental misunderstanding of the true location of vulnerability.
This represents the most severe type of analytical failure,
where 
AVR tools fix symptoms close to the point of failure
rather than addressing the underlying cause that
eventually leads the vulnerability.
To illustrate this failure mode, consider a NULL pointer dereference vulnerability in AST module of Python interpreter demonstrated by this PoC exploit:
\begin{minted}[fontsize=\footnotesize,breaklines]{python}
import ast; del ast.AST._fields; t = ast.AST(arg1=123)
\end{minted}
When \texttt{\small ast.AST.\_fields} is deleted and a new AST object is subsequently created with \texttt{\small ast.AST(arg1=123)}, the program crashes with a NULL pointer dereference.
\autoref{fig:incorrect_root_cause_patches} shows how the AVR tools
and developer approach this problem with fundamentally different philosophies.
%
The AVR-generated patch applies a band-aid solution by adding a defensive NULL check in \texttt{\small PySequence\_Contains()} where the crash occurs. While this mutes the crash, it fails to address why the NULL condition arose in the first place. In contrast, the developer's patch identifies the root cause in \texttt{\small ast\_type\_init()}, where the initialization code incorrectly uses \texttt{\small PyObject\_GetOptionalAttr()} to retrieve \texttt{\small \_fields}. By changing to \texttt{\small PyObject\_GetAttr()}, the patch enforces stricter validation during object creation.

The developer's approach is grounded in the AST specification, which explicitly states that
\emph{``Each concrete class has an attribute \texttt{\small \_fields} which gives the names of all child nodes''}~\cite{python_ast}. This makes \texttt{\small \_fields} a mandatory component of the AST object model, not an optional one.
The human patch recognizes this specification requirement and prevents invalid objects from being created, while the AVR patch merely handles the consequences of allowing such invalid objects to exist.
This example demonstrates how incorrect root cause identification leads to patches that may prevent immediate crashes but fail to address the underlying design violation, potentially leaving the system vulnerable to related issues. 


\subsection{Specification Violation}

We classify patches into this category
when the generated patch correctly identifies the problematic code location
but produces modifications that violate
established software specifications, 
programming language standards, or
documented functional requirements.
These violations typically manifest as changes to
input validation logic,
return value semantics, or
control flow that contradict to properties or contracts
communicated otherwise.
As demonstrated in \autoref{sec:motivation} with the PHP \texttt{\small range()} function, the AVR-generated patch enforced strict type checking by rejecting mixed-type inputs with runtime errors,
while the PHP language specification requires permissive type coercion that allows mixed numeric and string arguments to generate valid numeric ranges. 
This violation 
demonstrates how patches can on one hand fix vulnerabilities while
on the other hand fundamentally breaking expected program functionality.

Patches in specification violation category
demonstrate partial understanding of the vulnerability context
but fail to respect the intended functionality of the target system overall.
This failure mode indicates that
AVR systems need improved adherence to program specifications,
requiring better integration of 
a knowledge base that captures intended program behaviors,
and potential constraint validation during patch generation.


\begin{figure}[t]
\centering
\begin{subfigure}[t]{0.48\textwidth}
\centering
\begin{minted}[fontsize=\scriptsize,linenos,breaklines,numbersep=3pt,xleftmargin=1pt]{diff}
@@ -661,8 +661,12 @@ getArgumentAccessInfo(...) {
   auto TypeSize = DL.getTypeStoreSize(Ty);
   if (!TypeSize.isScalable() && Offset) {
     int64_t Size = TypeSize.getFixedValue();
-    return ConstantRange(APInt(64, *Offset, true),
-                         APInt(64, *Offset + Size, 
-                         true));
+    if (Size > 0) {
+      int64_t End = *Offset + Size;
+      if (End > *Offset)
+        return ConstantRange(APInt(64, *Offset, true),
+                             APInt(64, End, true));
+    }
   }
   return std::nullopt;
 };
\end{minted}
\caption{AVR-Generated Patch}
\label{fig:llm_patch}
\end{subfigure}
\hfill
\begin{subfigure}[t]{0.48\textwidth}
\centering
\begin{minted}[fontsize=\scriptsize,linenos,breaklines,numbersep=3pt,xleftmargin=1pt]{diff}
@@ -661,8 +661,13 @@ getArgumentAccessInfo(...) {
   auto TypeSize = DL.getTypeStoreSize(Ty);
   if (!TypeSize.isScalable() && Offset) {
     int64_t Size = TypeSize.getFixedValue();
-    return ConstantRange(APInt(64, *Offset, true),
-                         APInt(64, *Offset + Size, 
-                         true));
+    APInt Low(64, *Offset, true);
+    bool Overflow;
+    APInt High = Low.sadd_ov(APInt(64, Size, true), 
+                             Overflow);
+    // Bail if the range overflows signed 64-bit int.
+    if (Overflow)
+      return std::nullopt;
+    return ConstantRange(Low, High);
   }
   return std::nullopt;
 };
\end{minted}
\caption{Developer Patch}
\label{fig:human_patch}
\end{subfigure}
\Description{Domain Ignorance Example: Comparison of AVR and developer patches for an integer overflow bug in LLVM.}
\caption{Domain Ignorance Example}
\label{fig:llvm_patches}
\end{figure}

\subsection{Poor Code Practice}

We classify patches into this category
when the generated patch correctly identifies the vulnerability location and
maintains specification compliance
but still fails \testname due to
poor code quality or
violation of established coding practices.
These patches demonstrate adequate analytical capabilities of LLM but
reveal insufficient technical domain knowledge or
adherence to project-specific design principles.
This failure mode indicates that
AVR systems need enhanced code generation capabilities by
incorporating better understanding of
platform-specific requirements,
software engineering best practices, and sometimes even
intricacies in compiler behaviors.
We provide two examples to 
illustrate patches in this category.
One example involves an AVR-generated patch that
violates C/C++ programming standards (i.e., causing undefined behavior),
which is shown in \autoref{fig:llvm_patches}. The vulnerability occurs when computing memory access ranges in the
\texttt{\small FunctionAttrs()} optimization pass,
where \texttt{\small *Offset + Size} can overflow and wrap around.

The AVR patch in \autoref{fig:llm_patch} demonstrates overflow detection by adding a simple comparison \texttt{\small End > *Offset} to check for overflow. However, this approach is flawed because it relies on undefined behavior---when signed integer overflow occurs in C++, the comparison itself invokes undefined behavior, and aggressive compiler optimizations may eliminate the check entirely, assuming that signed overflow never occurs. The patch creates a false sense of security while potentially being optimized away at compile time.
In contrast, the developer patch in \autoref{fig:human_patch} correctly uses LLVM's \texttt{\small APInt::sadd\_ov} method, which is specifically designed for overflow-safe arithmetic operations. This approach uses well-defined overflow detection mechanisms that cannot be optimized away by compilers, properly handling the edge case by returning \texttt{\small std::nullopt} when overflow is detected. The developer solution demonstrates deep understanding of both the vulnerability context and the platform-specific requirements for reliable overflow detection in optimized code.

\begin{figure}[t]
    \centering
\begin{minted}[fontsize=\footnotesize,linenos,breaklines,numbersep=5pt,xleftmargin=1pt]{diff}
   GET_NODE(sxe, node);
+  if (!node) {
+    /* avoid null dereference */
+    return &EG(err_zval);
+  }
   php_libxml_invalidate_node_from_doc(node->doc); 
   if (node) {
        if (attribs) {
\end{minted}

\Description{Logic Shortcuts Example: A patch that adds an early null check and immediate return, making the subsequent if (node) conditional redundant.}
\caption{Logic Shortcuts Example}
\label{fig:simplexml_fixes}
\end{figure}

The other example demonstrates how a AVR-generated patch
disregards the control flow logic carefully maintained by developers. This example encompasses patches that introduce logically destructive code structures that break the developer's intended design patterns.
%
As illustrated in \autoref{fig:simplexml_fixes}, the generated patch adds an early null check that immediately returns when \texttt{\small node} is null, bypassing the developer's carefully structured conditional logic that was designed to handle null cases gracefully within the existing control flow
(the matching \texttt{\small else} block for the \texttt{\small if} condition at line 7).
This approach creates destructive code in the later \texttt{\small if(node)} check, as the condition will always be true since null values have already been filtered out by the early return, and also violates the developer's intent to maintain a unified error handling strategy throughout the function. The correct approach respects the original design by moving the \texttt{\small php\_libxml\_invalidate\_node\_from\_doc} call inside the existing \texttt{\small if(node)} conditional, preserving the developer's intended control flow while eliminating the vulnerability without introducing
structural inconsistencies.

%% file: sections/8_implication.tex
\section{Implications for AVR Research}

Our evaluation on \dataset reveals that conventional test suite-based validation methods may substantially overestimate the effectiveness of AVR tools, with over 40\% of patches that pass basic tests failing to pass \testname tests. Our further analysis of these false positive cases suggests that the primary cause is specification violations, where patches produce modifications that violate established software specifications or documented behavioral requirements. These findings suggest two main implications for AVR research.

\textbf{AVR research could benefit from adopting more reliable validation methodologies during evaluation beyond simply running PoC exploits and existing functional tests.} The discovery of such high false discovery rates (40\%+) across all tested state-of-the-art AVR systems indicates that current validation practices may create an illusion of effectiveness. This systematic overestimation could undermine confidence in deployment decisions and suggests that some published success rates might be inflated. To address this potential gap, future AVR evaluation frameworks could incorporate multi-layered validation approaches, including developer-authored functional tests like \testname tests, manual comparison against developer patches to assess semantic equivalence and formal verification techniques. Research might prioritize developing automated methods to generate or identify comprehensive test suites that capture true functional requirements, while AVR benchmarks could consider including more rigorous validation criteria that help ensure patches meet both security and specification requirements for potential production deployment.

\textbf{Current AVR approaches that rely primarily on codebase and vulnerability information as input may be insufficient for generating production-ready patches.} Our categorical analysis in \autoref{sec:finding} reveals that modern LLM-based AVR tools often struggle to understand program specifications, API semantics, and behavioral requirements that appear essential for correct repairs. The prevalence of specification violations suggests that many critical requirements are documented in external sources rather than being easily inferable from code alone. AVR systems could benefit from incorporating information from project documentation, API specifications, coding guidelines, comments, and other textual resources that capture intended program behavior and constraints. Future AVR research might explore methods to automatically extract and leverage such documented requirements, develop techniques to integrate natural language specifications with code analysis, and create approaches that can reason about both explicit code patterns and implicit behavioral expectations described in documentation. This potential shift toward incorporating documented knowledge alongside code analysis could represent an important evolution for generating patches that respect both syntactic correctness and documented program intentions.

%% file: sections/9_discussion.tex
\section{Discussion}

\noindent\textbf{Scope and Limitations.} 
While our study provides insights into the limitations of conventional test suite-based patch validation, several constraints limit the generalizability of our findings. First, \dataset focuses exclusively on C/C++ programs across 20 open-source projects, which may not represent the full spectrum of programming languages and software domains where AVR tools are applied. Different languages have varying 
features (e.g., type systems) and testing cultures that could influence both vulnerability patterns and validation effectiveness. Second, improved patch validation methods do not directly enhance the effectiveness of AVR tools. Although some prior work suggests that better patch validation can provide more informative feedback to improve AVR systems, our test generation approach still relies on developer-provided correct patches rather than generating tests from scratch. Consequently, our method is primarily suited for evaluation purposes rather than integration into an end-to-end repair pipeline. We leave the generation of \testname tests from scratch for future work.


\noindent\textbf{Automated Software Engineering.} Building \dataset required substantial effort due to challenges such as resolving complex dependencies and aggregating information from multiple sources. This labor-intensive process consumed considerable time, limiting both dataset scale and diversity. Our experience highlights the critical need for automated software engineering.
Recent advances have begun to address these challenges: CompileAgent\cite{hu2025compileagent} automates repository-level compilation with LLM agents, while ExecutionAgent\cite{bouzenia2025you} can automatically executes functional test suites. Automated dataset construction systems like SWE-smith\cite{yang2025swe} and SWE-rebench\cite{badertdinov2025swe} have demonstrated the ability to create datasets an order of magnitude larger than manual collections.


%% file: sections/10_related.tex
\section{Related Work}

\noindent\textbf{Patch Equivalence.}
Among prior works that use test suite-based methods
to evaluate the patch correctness rate of their tools,
some \cite{SAN2PATCH,pearce2023examining} also employ manual comparison to count the number of generated patches that are semantically equivalent to developer patches---%
establishing a lower bound of
the actual patch success rate.
Empirical studies have shown that approximately 25\% of correct AVR patches are syntactically different but semantically equivalent to developer patches \cite{wang2019different}.
Recent advances in automated semantic equivalence assessment combine syntactic and semantic similarity metrics with test coverage analysis \cite{ghanbari2022shibboleth}, while sophisticated approaches leverage program invariants and pre-trained language models for semantic reasoning about patch equivalence \cite{le2023invalidator}.
Research on program equivalence for adaptive vulnerability repair has formalized the patch equivalence problem using test-equivalence relations, proposing efficient algorithms for partitioning patches into equivalence classes based on runtime behavior \cite{weimer2013leveraging}. 
Our work provides a better approach to calculate
a tighter upper bound of patch success rate,
while the lower bound is still defined by equivalence.

\noindent\textbf{Formal Verification.} Formal verification approaches offer mathematically rigorous alternatives to test-suite based patch validation, providing stronger guarantees about patch correctness through specification-based reasoning and constraint solving. SemFix \cite{nguyen2013semfix} formulates repair requirements as constraints solved by SMT solvers, providing formal guarantees about patch correctness beyond test adequacy. Subsequent advances in scalable semantics-based repair using symbolic execution with Z3 SMT solver have demonstrated practical applications of formal methods to multiline vulnerability repair \cite{mechtaev2016angelix}. Contract-based repair approaches use formal specifications such as pre- and postconditions for both patch generation and validation \cite{wei2010automated}, while sound and complete mutation-based vulnerability repair provides theoretical guarantees of soundness and completeness through bounded model 
and SAT/SMT 
\cite{rothenberg2016sound}. 
Advanced approaches integrate formal verification throughout the repair process, using constraint solving with mathematical foundations such as Farkas lemma for 
patch synthesis \cite{nguyen2019automatic}, while modular program verifiers \cite{logozzo2012modular} enable property-specific validation.


%% file: sections/11_conclusion.tex
\section{Conclusion}

Our study suggests that current AVR evaluation practices may benefit from more comprehensive validation approaches, as we observe that over 40\% of patches deemed correct by conventional test-suite methods fail when evaluated against \testname tests across three state-of-the-art LLM-based AVR systems. Our proposed \testname test demonstrates promising results, with over 70\% of passing patches achieving semantic equivalence with developer solutions, suggesting this methodology could serve as a valuable complement to existing validation practices. These findings encourage future research to consider incorporating more rigorous quality assessment approaches and including specification information in AVR.

%% file: sections/appendix.tex
\noindent
\begin{minipage}{\textwidth}
\centering
\includegraphics[width=\linewidth]{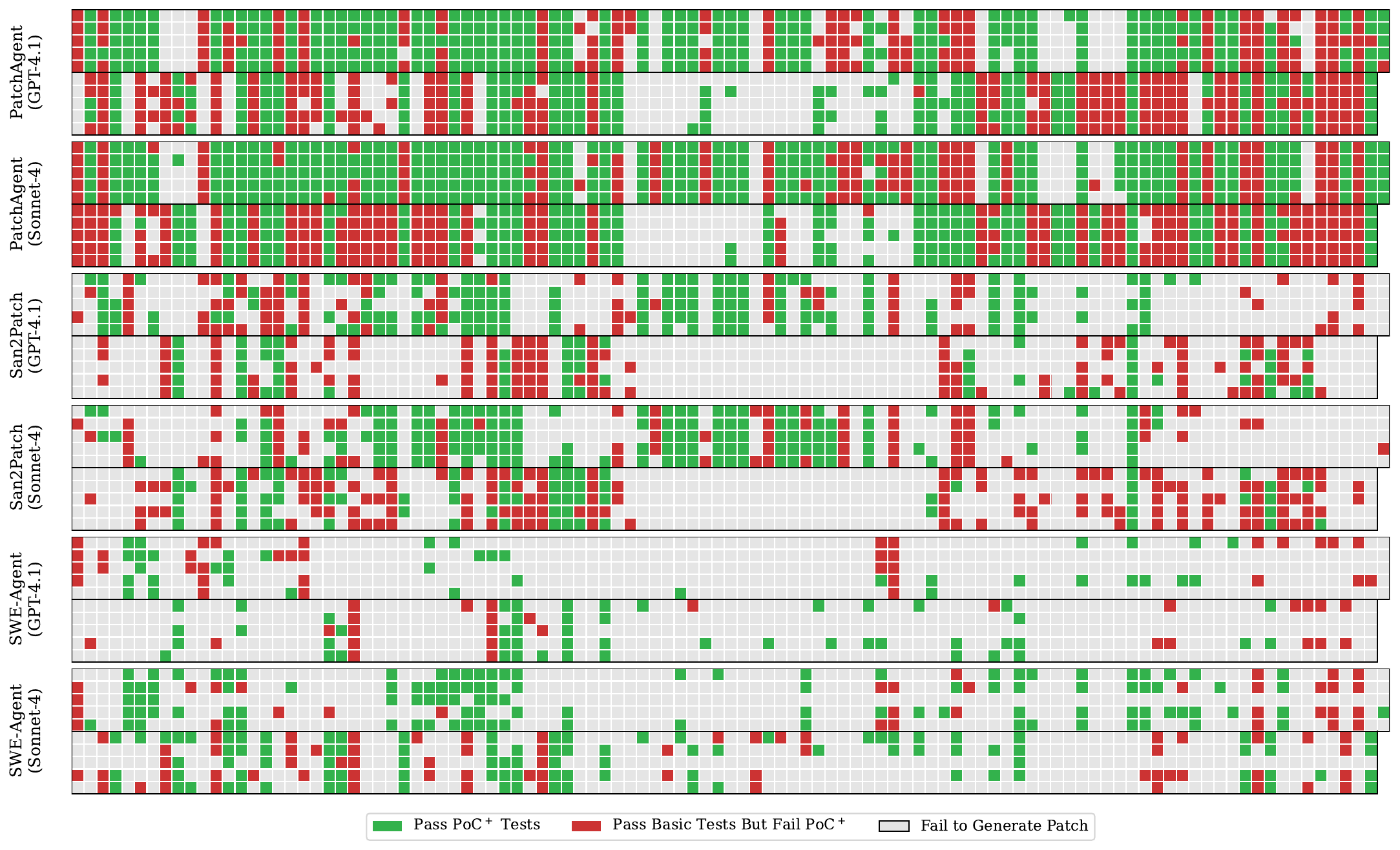}
\captionof{figure}{Comprehensive evaluation results across 209 vulnerability cases.}
\label{fig:full_results}
\vspace{3em}

\noindent
\begin{minipage}[t]{0.48\textwidth}

\appendix
\section{Appendix}

\autoref{tab:auto_gen} presents the paths to automated test generation scripts for projects that natively support \testname generation, as evaluated in the \textit{Output Checking} category.
\autoref{fig:full_results} presents the comprehensive results of our evaluation across all 209 vulnerability cases. The visualization employs a grid-based format where each agent-model combination is represented by a 10$\times$105 matrix. Due to the odd number of vulnerability cases (209), the data is split into two unequal halves: the top 5 rows display results for vulnerabilities 1-105, while the bottom 5 rows show results for vulnerabilities 106-209.
Each column in the grid corresponds to a single vulnerability case, with 5 cells per column representing independent experimental runs. The color coding follows our evaluation criteria: green cells indicate successful patches that pass both basic functionality tests and \testname tests, red cells represent patches that pass basic tests but fail the more stringent \testname requirements, and gray cells denote cases where the agent failed to generate any viable patch.

\section{Open Science.}\label{sec:open_science}

To support the evaluation of this paper's contributions and facilitate reproducibility, we provide all artifacts at \url{https://github.com/cla7aye15I4nd/PVBench}. The repository contains the \dataset dataset and the original experimental data from evaluating three state-of-the-art agents on \dataset.

\end{minipage}%
\hfill
\begin{minipage}[t]{0.50\textwidth}
\centering
\small
\captionof{table}{Script Path of Projects that Support \testname Generation}
\label{tab:auto_gen}
\begin{tabular}{ll}
\toprule
\textbf{Project} & \textbf{Automated Test Generation Scripts Path} \\
\midrule
Hermes  & \href{https://github.com/facebook/hermes/blob/main/utils/updateErrorTest.py}{\texttt{utils/updateErrorTest.py}} \\
libxml2 & 
        \begin{tabular}[t]{@{}l@{}}
            \href{https://github.com/GNOME/libxml2/blob/master/codegen/genTestApi.py}
            {\texttt{codegen/genTestApi.py}} \\
            \href{https://github.com/GNOME/libxml2/blob/master/xstc/fixup-tests.py}
            {\texttt{xstc/fixup-tests.py}} \\
        \end{tabular} \\
LLVM    & \begin{tabular}[t]{@{}l@{}}
            \href{https://github.com/llvm/llvm-project/blob/main/llvm/utils/update_analyze_test_checks.py}
            {\texttt{llvm/utils/update\_analyze\_test\_checks.py}} \\
            \href{https://github.com/llvm/llvm-project/blob/main/llvm/utils/update_any_test_checks.py}
            {\texttt{llvm/utils/update\_any\_test\_checks.py}} \\
            \href{https://github.com/llvm/llvm-project/blob/main/llvm/utils/update_cc_test_checks.py}
            {\texttt{llvm/utils/update\_cc\_test\_checks.py}} \\
            \href{https://github.com/llvm/llvm-project/blob/main/llvm/utils/update_llc_test_checks.py}
            {\texttt{llvm/utils/update\_llc\_test\_checks.py}} \\
            \href{https://github.com/llvm/llvm-project/blob/main/llvm/utils/update_mca_test_checks.py}
            {\texttt{llvm/utils/update\_mca\_test\_checks.py}} \\
            \href{https://github.com/llvm/llvm-project/blob/main/llvm/utils/update_mir_test_checks.py}
            {\texttt{llvm/utils/update\_mir\_test\_checks.py}} \\
            \href{https://github.com/llvm/llvm-project/blob/main/llvm/utils/update_test_checks.py}
            {\texttt{llvm/utils/update\_test\_checks.py}} \\
            \href{https://github.com/llvm/llvm-project/blob/main/llvm/utils/update_test_prefix.py}
            {\texttt{llvm/utils/update\_test\_prefix.py}}
          \end{tabular} \\
PHP     & \href{https://github.com/php/php-src/blob/master/scripts/dev/bless_tests.php}{\texttt{scripts/dev/bless\_tests.php}} \\
Wabt    & 
\begin{tabular}[t]{@{}l@{}}
    \href{https://github.com/WebAssembly/wabt/blob/main/test/run-tests.py}
{\texttt{test/run-tests.py}} \\
    \href{https://github.com/WebAssembly/wabt/blob/main/test/update-spec-tests.py}
{\texttt{test/update-spec-tests.py}} \\
\end{tabular} \\
\bottomrule
\end{tabular}
\end{minipage}
\end{minipage}